\documentclass[english]{article}
\usepackage[T1]{fontenc}
\usepackage[latin1]{inputenc}
\usepackage{geometry}
\usepackage{float}
\usepackage{mathrsfs}
\usepackage{amsmath}
\usepackage{amssymb}
\usepackage{graphicx}
\usepackage{subfigure}
\usepackage[textsize=tiny]{todonotes}

\makeatletter

\floatstyle{ruled}
\newfloat{algorithm}{tbp}{loa}
\providecommand{\algorithmname}{Algorithm}
\floatname{algorithm}{\protect\algorithmname}

\usepackage{amsthm}

\usepackage{mathrsfs}

\usepackage{amsfonts}

\usepackage{epsfig}

\usepackage{mathrsfs}

\usepackage{enumerate}

\newcommand{\bbR}{\mathbb R}

\newtheorem{ass}{Assumption}[section]
\newcounter{hypA}

\newcommand{\cG}{\mathcal{G}}

\newcommand{\hu}{\hat {u}}

\newcommand{\bbE}{\mathbb{E}}
\newcommand{\bbP}{\mathbb{P}}

\newcommand{\cO}{\mathcal{O}}

\usepackage{xspace}
\usepackage{tabu}
\usepackage{booktabs}

\def\bbR{\mathbb{R}}

\usepackage{babel}\date{}

\usepackage{babel}

\makeatother

\usepackage{babel}
\begin{document}

\begin{center}

{\Large \textbf{Advanced Multilevel Monte Carlo Methods}}

\vspace{0.5cm}

BY AJAY JASRA$^{1}$, KODY LAW$^{2}$, \& CARINA SUCIU$^{3}$ 

{\footnotesize $^{1}$Department of Statistics \& Applied Probability,
National University of Singapore, Singapore, 117546, SG.}
{\footnotesize E-Mail:\,}\texttt{\emph{\footnotesize staja@nus.edu.sg}}\\
{\footnotesize $^{2}$Computer Science and Mathematics Division,
Oak Ridge National Laboratory,  Oak Ridge, 37934 TN, USA.}
{\footnotesize E-Mail:\,}\texttt{\emph{\footnotesize kodylaw@gmail.com}}\\
{\footnotesize $^{3}$Center for Uncertainty Quantification
in Computational Science \& Engineering, King Abdullah University of Science and Technology, Thuwal, 23955-6900, KSA.}\\
{\footnotesize E-Mail:\,}\texttt{\emph{\footnotesize oana.suciu@kaust.edu.sa}}
\end{center}

\begin{abstract}
\noindent This article reviews the application of
advanced Monte Carlo techniques in the context of 
 Multilevel Monte Carlo (MLMC). MLMC 
is a strategy employed to compute expectations which can be biased in some sense, for instance, by using the discretization of a associated probability law. 
The MLMC approach works with a hierarchy of biased approximations
which become progressively more accurate and more expensive. 
Using a telescoping representation of the most accurate approximation, 
the method is able to reduce the computational cost for a given level of error versus i.i.d. sampling from this latter approximation. 
All of these ideas 
originated for cases where exact sampling from couples in the hierarchy is possible. 
This article considers the case where such exact sampling is not currently possible. 
We 
consider Markov chain Monte Carlo and sequential Monte Carlo 
methods which have been 
introduced in the literature and we describe 
different strategies which facilitate the application of MLMC within these methods.
\end{abstract}

\section{Introduction}

Let $(\mathsf{E},\mathcal{E})$ be a measurable space, $\pi$ be a probability measure on $(\mathsf{E},\mathcal{E})$
and $\varphi:\mathsf{E}\rightarrow\mathbb{R}$ be a measurable and $\pi-$integrable function. In this article we are concerned with the computation of
\begin{equation}\label{eq:basic_exp}
\mathbb{E}_{\pi}[\varphi(U)] = \int_{\mathsf{E}} \varphi(u) \pi(du)
\end{equation}
for many different $\pi-$integrable functions $\varphi$. In addition, if $\pi$ admits a density
w.r.t.~a dominating $\sigma-$finite measure $du$ and if one can write for $\kappa:\mathsf{E}\rightarrow\mathbb{R}_+$
\begin{equation}\label{eq:ratio}
\pi(du) = \frac{\kappa(u)}{Z}du,
\end{equation}
where $Z$ is not known, but one can obtain $\kappa$ up-to a non-negative unbiased estimator, then
one is also interested in the computation of $Z$. These problems occur in a wide variety of real applications,
often concerning Bayesian statistical inference. For instance, the computation of \eqref{eq:basic_exp} 
can be associated to posterior expectations, and the value of $Z$ can be used for Bayesian model 
selection; see ~\cite{robert}. Many of these problems are found in many real applications, such as meteorology,
finance and engineering; see \cite{law2015data,robert}. Later in this article,
we will expand upon the basic problem here.

We focus on the case when $\pi$ is associated to some complex 
continuum problem, for instance, a
continuous-time stochastic process or the solution of a partial differential equation (PDE), although the methodology
described in this article is not constrained to such examples. 
Also, we will assume that:
\begin{enumerate}
\item{One must resort to numerical methods to approximate \eqref{eq:basic_exp} or $Z$.}
\item{One can, at best, hope to approximate expectations w.r.t.~some \emph{biased} version
of $\pi$, call it $\pi_L$. It is explicitly assumed that this bias is associated with a \emph{scalar} parameter $h_L\in\mathbb{R}_+$ and that the bias disappears as $h_L\rightarrow 0$.}
\item{When using Monte Carlo methods, exact sampling from $\pi_L$ is not possible; that is, one cannot sample i.i.d.~from $\pi_L$.}
\end {enumerate}
Examples of models satisfying 1.~\& 2.~include laws of stochastic differential equations (SDE) 
for which one
cannot sample exactly (e.g., ~\cite{KlPl92}), and one resorts to Euler or Milstein discretization. In addition, the law of a quantity of interest (QOI) resulting from the solution of a PDE 
associated to random input parameters, which cannot be solved exactly and needs to be numerically approximated. 
Examples satisfying 1.-3.~include for example Bayesian instances of the above,
where one updates the prior probability distribution based on noisy data to obtain 
the posterior conditional on the observed data
(e.g., ~\cite{hoan:12}) or 
general models where approximate Bayesian computation 
(e.g., ~\cite{marin}) must be used.

\subsection{Monte Carlo and Multilevel Monte Carlo}\label{sec:ml_mlmc}

For now, let us assume that only 1.~\& 2.~apply. 
In this context, one could, for instance, sample $U^1,\dots,U^N$ i.i.d.~from
$\pi_L$ and one could use the standard Monte Carlo estimator 
$\frac{1}{N}\sum_{i=1}^N \varphi(u^i)$, which approximates
$\mathbb{E}_{\pi_L}[\varphi(U)]=\int_{\mathsf{E}}\varphi(u)\pi_L(du)$. 
To explain what will follow, we will suppose the following.

\begin{ass}[Cost and discretization error] There are $\alpha, \zeta>0$ such that 
\begin{itemize}
\item{The cost of simulating one sample is $\cO(h_L^{-\zeta})$.}
\item{The bias is of order $\mathcal{O}(h_L^{\alpha})$.}
\end{itemize}
\label{ass:bias}
\end{ass}

This scenario would occur 
under an Euler discretization of a suitably regular SDE 
and for appropriate $\varphi$, with $\alpha=\zeta=1$.
For simplicity suppose, $h_L=2^{-L}$. 
Consider the mean square error (MSE) associated to the Monte Carlo estimator
$$
\mathbb{E}\Big[\Big(\frac{1}{N}\sum_{i=1}^N \varphi(U^i)-\mathbb{E}_{\pi}[\varphi(U)]\Big)^2\Big],
$$
where $\mathbb{E}[\cdot]$ is the expectation operator w.r.t.~the 
distribution of the samples $(U^1,\dots,U^N)$. 
Note that $\bbE[\varphi(U^i)]=\mathbb{E}_{\pi_L}[\varphi(U)]$.
Adding and
subtracting $\mathbb{E}_{\pi_L}[\varphi(U)]$, 
and assuming $\varphi$ has a second moment w.r.t.~$\pi_L$,
one has that the MSE is equal to 
$$
\frac{1}{N}\mathbb{V}\textrm{ar}_{\pi_L}[\varphi(U)] + (\mathbb{E}_{\pi_L}[\varphi(U)]-\mathbb{E}_{\pi}[\varphi(U)])^2,
$$
which is the standard variance plus bias squared. 
Now let $1>\epsilon>0$ be given, and suppose one wants to control
the MSE, e.g.,~so that it is $\mathcal{O}(\epsilon^2)$. 
One begin by controlling the bias, by setting $L$. 
The constraint that $2^{-2\alpha L} = \cO(\epsilon^2)$ 
can be satisfied by choosing 
$L \propto -\frac{\log(\epsilon)}{\alpha\log(2)}$.
Then the constraint that the variance is $\mathcal{O}(\epsilon^{2})$
can be satisfied by choosing $N \propto \epsilon^{-2}$.
The cost can then be controlled by 
$2^{\zeta L}\epsilon^{-2}=
\cO(\epsilon^{-2-\frac{\zeta}{\alpha}})$. 
In the case of an Euler discretization of a sufficiently regular SDE, 
one can asymptotically obtain an MSE of $\mathcal{O}(\epsilon^2)$
for a cost of $\mathcal{O}(\epsilon^{-3})$.

The multilevel Monte Carlo (MLMC) method \cite{giles,giles1,hein} 
is designed to improve over the cost of Monte Carlo. 
As above, suppose $h_l=2^{-l}$.
The idea is to consider
a hierarchy, $\infty>h_1>\cdots>h_L>0$
and consider the respresentation
\begin{equation}\label{eq:ml_approx}
\mathbb{E}_{\pi_L}[\varphi(U)] = \sum_{l=1}^L \{\mathbb{E}_{\pi_l}-\mathbb{E}_{\pi_{l-1}}\}[\varphi(U)],
\end{equation}
where for $1\leq l\leq L$, $\mathbb{E}_{\pi_l}$ is the expectation w.r.t.~$\pi_l$ (i.e.,~the biased approximation with parameter $h_l$)
and for $l=1$, $\mathbb{E}_{\pi_{l-1}}[\varphi(U)]:=0$. 
This will be referred to in what follows as the ML identity.
Here, it is assumed that for each probability $\pi_l$ one is only interested
in a marginal on $\mathsf{E}$, even if the entire space must be enlarged to facilitate the biased approximation. So, for instance,
$\pi_l$ may be defined on a larger 
space than $\mathsf{E}$, but it admits a marginal on $\mathsf{E}$ which approaches $\pi$ as
$l$ grows. Furthermore, it is explicitly assumed that the cost of sampling or evaluating 
$\pi_l$ grows with $l$; again this would occur in most applications mentioned above.
To approximate the first term in the summation of \eqref{eq:ml_approx}, one samples $U^1(1),\dots,U^{N_1}(1)$ i.i.d.~from
$\pi_1$ and one uses the standard Monte Carlo estimator $\frac{1}{N}\sum_{i=1}^N \varphi(u^i(1))$. 
For the remainder of the terms $2\leq l\leq L$, we suppose that it is possible to sample a (dependent) \emph{coupling} of $(\pi_l,\pi_{l-1})$ with samples $(U_l(l),U_{l-1}(l))$
such that the following holds.

\begin{ass}[Variance] There is a $\beta>0$ such that 
the variance w.r.t.~the coupling of $(\pi_l,\pi_{l-1})$, 
$$\mathbb{V}\textrm{ar}_{(\pi_l,\pi_{l-1})}[\varphi(U_l(l))-\varphi(U_{l-1}(l))]=\mathcal{O}(h_l^{\beta}).$$
\label{ass:var}
\end{ass}

Note that by coupling, we mean that $U_l(l)\sim\pi_l$ and $U_{l-1}(l)\sim\pi_{l-1}$ (the random variables are generally dependent). 
{Couplings which satisfy the above bullet exist, for instance, in the context of SDE.
If the SDE is suitably regular and is approximated by Euler method, 
then $\beta=1$, or $\beta=2$ if the diffusion coefficient is constant.}
In order to approximate the summands in \eqref{eq:ml_approx} for $2\leq l\leq L$, 
draw $N_l$ 
i.i.d.~samples $(U_l^1(l),U_{l-1}^1(l)),\dots,(U_l^{N_l}(l),U_{l-1}^{N_l}(l))$ 
from the coupling $(\pi_l,\pi_{l-1})$,
and use the unbiased estimator
$$
\frac{1}{N_l}\sum_{i=1}^{N_l}\{\varphi(u_l^i(l))-\varphi(u_{l-1}^i(l))\} 
\approx \{\mathbb{E}_{\pi_l}-\mathbb{E}_{\pi_{l-1}}\}[\varphi(U)].
$$

The multilevel estimator is thus
$$
\frac{1}{N_1}\sum_{i=1}^{N_1}\varphi(u_1^i(1)) + \sum_{l=2}^L\Big(\frac{1}{N_l}\sum_{i=1}^{N_l}\{\varphi(u_l^i(l))-\varphi(u_{l-1}^i(l))\}\Big).
$$
One can analyze the MSE as above. 
It is equal to 
$$
\frac{1}{N_1}\mathbb{V}\textrm{ar}_{\pi_1}[\varphi(U)]
+\sum_{l=2}^L \frac{1}{N_l} \mathbb{V}\textrm{ar}_{(\pi_l,\pi_{l-1})}[\varphi(U_l(l))-\varphi(U_{l-1}(l))]
+ (\mathbb{E}_{\pi_L}[\varphi(U)]-\mathbb{E}_{\pi}[\varphi(U)])^2 \, ,
$$
and the associated cost is $\sum_{l=1} N_l h_l^{-\zeta}$, where we assume that the cost of sampling the coupling $(\pi_l,\pi_{l-1})$
is at most the cost of sampling $\pi_l$. 
Since we have assumed $h_1=\mathcal{O}(1)$, then
$\frac{1}{N_1}\mathbb{V}\textrm{ar}_{\pi_1}[\varphi(U)]\leq \frac{C}{N_1}$,
for $\infty>C>0$ a constant independent of $l$.
Now let $1>\epsilon>0$ be given, and suppose one wants to control
the MSE, e.g., ~so that it is $\mathcal{O}(\epsilon^2)$. 
One controls the bias as above by 
letting 
\begin{equation}\label{eq:biasL}
L\propto-\frac{\log(\epsilon)}{\alpha\log(2)}.
\end{equation} 
Then one seeks to minimize the cost $\sum_{l=1} N_l h_l^{-\zeta}$
in terms of $N_1,\dots,N_L$, 
subject to the constraint 
$$
\sum_{l=1}^L \frac{h_l^{\beta}}{N_l} \propto \epsilon^2. 
$$
This constrained optimization problem is solved in \cite{giles} and has the solution 
$N_l\propto h_l^{(\beta+\zeta)/2}$ to obtain a MSE of
$\mathcal{O}(\epsilon^2)$.
Solving for the Lagrange multiplier, with equality above, 
one has that 
\begin{equation}\label{eq:nell}
N_l=\epsilon^{-2} h_l^{(\beta+\zeta)/2}K_L \, ,
\end{equation} 
where $K_L=\sum_{l=1}^L h_l^{(\beta-\zeta)/2}$.
Note that $K_L$ 
may depend upon $L$, depending upon the values of $\beta,\zeta$. 
In the Euler case $\beta=\zeta$. 
So, one is able to obtain an MSE of $\mathcal{O}(\epsilon^2)$
for the cost $\mathcal{O}(\epsilon^{-2}\log(\epsilon)^2)$.  
In the special case in which the diffusion coefficient is constant, one 
obtains the Milstein method with $\beta>\zeta$, so the cost can be controlled by 
$\mathcal{O}(\epsilon^{-2})$.

The MLMC framework discussed above is considered  
in various different guises in this paper.
The cost will always scale as in Assumption \ref{ass:bias}, and some 
analogue of the bias from Assumption \ref{ass:bias} will determine 
$L$ as defined in \eqref{eq:biasL}.
We will then require rates on different quantities analogous to 
Assumption \ref{ass:var}, in order to ensure the choice of $N_l$
in \eqref{eq:nell} is optimal.

\subsection{Methodology Reviewed}

In the above section, we have supposed only the points 1. and  2. 
However, in this article we are considering all three points.
In other words, it is not possible to exactly sample from any of 
$\pi_L$ 
or $\pi_1,\dots,\pi_{L-1}$. 
One of the critical
ingredients of the MLMC method is 
sampling dependent couples of the pairs $(\pi_l,\pi_{l-1})$, which one
might argue is even more challenging than sampling from $\pi_L$ for a single given $L$.  
In the context of interest, 
one might use Markov chain Monte Carlo (MCMC -- see e.g.,~\cite{robert,roberts}) 
or Sequential Monte Carlo (SMC -- see e.g.,~\cite{delm:04,delmoral1,doucet_johan, doucet_def}) 
to overcome the challenges of not being able to sample $\pi_L$. 
However, a simple procedure of trying
to approximate the ML identify \eqref{eq:ml_approx} by sampling independent MCMC chains targeting 
$\pi_1,\dots,\pi_{L}$ would seldom lead to improvement over just sampling from $\pi_L$. 
So, in such contexts where also using the MLMC approach makes sense, 
the main issue is how can one utilize such methodology so that 
one reduces the cost relative to 
exact sampling from $\pi_L$, for a given MSE.
There have been
many works on this topic and the objective of this article is to review these ideas as well as to identify important areas
which could be investigated in the future. 

The challenge 
lies not only in the design and application of the method, 
but in the subsequent analysis of the method, i.e.,~verifying that indeed it yields an improvement in cost for a given level of MSE. 
For instance, the analysis
of MCMC and SMC 
rely upon techniques in Markov chains (e.g.,~\cite{meyn,roberts}) 
and Feynman-Kac formulae (e.g.,~\cite{delm:04,delmoral1}). 
We highlight these techniques during our review.

\subsection{Structure}

This article is structured as follows. In Section \ref{sec:mot_ex}, we give a collection
of motivating examples from applied mathematics and statistics, for which the application
of multilevel methods would make sense, and are of interest from a practical perspective.
These examples are sufficiently complex that standard independent sampling
is not currently possible, but advanced simulation methods such as those described in the
previous subsection can be used.
In Section \ref{sec:comp_meth}, a short review of some of the computational methods
for which this review is focussed on is given. In Section \ref{sec:approach}, we review several
methods which have been adopted in the literature to date, mentioning the 
benefits and drawbacks of 
each approach. In Section \ref{sec:future}, some discussion of the potential for future work
is provided.

We end this introduction by mentioning that this review is not intended to be comprehensive. For
instance, we do not discuss quasi-Monte Carlo methods
or debiasing methods (e.g.,~\cite{rg:15}). An effort is of course made to discuss
as much work as possible that exists under the umbrella of 
advanced MLMC 
methods.  

\section{Motivating Examples}\label{sec:mot_ex}

\subsection{Bayesian Inverse Problems}\label{sec:bip}

We consider the following example as it is described in \cite{beskos} (see also \cite{hoan:12} and the references therein).

We introduce the 
 nested spaces $V := H^{1}(\Omega) \subset L^2(\Omega) \subset H^{-1} (\Omega)=: V^*$, 
where the domain $\Omega$ will be defined later.  
Furthermore, denote by $\langle \cdot, \cdot \rangle, \|\cdot\|$ the inner product and norm 
on $L^2$, 
and by $\langle \cdot, \cdot \rangle, |\cdot|$ 
the finite dimensional Euclidean inner product and norms. 
Denote weighted norms by adding a subscript as  
$\langle \cdot, \cdot \rangle_A := \langle A^{-\frac12}\cdot, A^{-\frac12}\cdot \rangle$, with corresponding norms
$|\cdot |_A$ or $\|\cdot \|_A$ for Euclidean and $L^2$ spaces, respectively 
(for symmetric, positive definite $A$ with $A^\frac12$ being the unique symmetric square root).

Let $\Omega \subset \bbR^D$ with $\partial \Omega \in C^1$ convex.
For $f \in V^*$, consider the following PDE on $\Omega$:
\begin{align}
\label{eq:uniellip}
- \nabla \cdot ( \hu \nabla p )  & =  f\  \quad  {\rm in} ~ \Omega\ , \\
p & = 0\  \quad  {\rm on} ~ \partial \Omega\ ,
\label{eq:bv}
\end{align}
where
\begin{equation}
\label{eq:expand}
\hu (x) = \bar{u}(x) + \sum_{k=1}^K u_k \sigma_k \phi_k(x) \ . 
\end{equation}
Define $u=\{u_k\}_{k=1}^K$, with $u_k \sim 
\mathcal{U}[-1,1]$ i.i.d. (the uniform distribution on $[-1,1]$). This determines the prior distribution for $u$.
  Assume that $\bar{u}, \phi_k \in C^\infty$ for all $k$ and that  
$\|\phi_k\|_\infty =1$.  
 In particular, 
 assume $\{\sigma_k\}_{k=1}^K$ 
 decay with $k$.  
The state space is $\mathsf{E} = \prod_{k=1}^K [-1,1]$.  
Assume the following property holds:
$\inf_x \hu(x) \geq  \inf_x \bar{u}(x) - \sum_{k=1}^K \sigma_k \geq u_* > 0$
so that the operator on the 
left-hand side of \eqref{eq:uniellip} is uniformly elliptic.  
Let $p(\cdot;u)$ denote the weak solution of \eqref{eq:uniellip} for parameter value $u$.  
Define the following vector-valued function 
$$
\cG(p) = [ g_1( p), \cdots , g_M ( p ) ]^\top\ , 
$$
where $g_m$ are elements of the dual space 
$V^*$ for  $m=1,\ldots, M$.
It is  assumed that the data take the form
\begin{equation}
Y = \cG (p) + \xi\ , \quad \xi \sim \mathcal{N}(0,\Gamma)\ , \quad \xi \perp u\ , 
\label{eq:data}
\end{equation}
where $\mathcal{N}(0,\Gamma)$ denotes the Gaussian random variable with mean $0$ and covariance $\Gamma$, 
and $\perp$ denotes independence. 
The unnormalized density for $u\in \mathsf{E}$ is then is given by:
\begin{equation*}
\kappa(u) = e^{-\Phi[\cG(p(\cdot;u))]}  \ , \quad \Phi(\cG) = \tfrac{1}{2}\, | \cG - y|^2_\Gamma
\ . 
\end{equation*}

\subsubsection{Approximation}

Consider the triangulated domains (with sufficiently regular triangles)
$\{\Omega^l\}_{l=1}^\infty$ approximating $\Omega$, 
where $l$ indexes the number of nodes $d_l\propto h_l^{-D}$, for 
triangulation diameter $h_l$, so that  we have 
$\Omega^1 \subset\cdots \subset  \Omega^{l} \subset \Omega^\infty :=\Omega$.
Furthermore, consider a finite element discretization on $\Omega^l$ 
consisting of $H^{1}$ functions $\{\psi_\ell\}_{\ell=1}^{d_l}$.
Denote the corresponding space
of functions of the form $\varphi = \sum_{\ell=1}^{d_l} v_\ell \psi^l_\ell$ by $V^l$, and notice that 
$V^1\subset V^{2}\subset \cdots \subset V^l \subset V$.  
By making the further Assumption 7 of 
\cite{hoan:12} that the weak solution $p(\cdot;u)$ of \eqref{eq:uniellip}-(\ref{eq:bv}) for parameter value $u$ 
is in the space $W=H^2 \cap H^1_0 \subset V$, one obtains a well-defined  
finite element approximation $p^l(\cdot;u)$ of $p(\cdot;u)$, 
with a rate of convergence in $V$ or $L^2$, independently of $u$.
Thus, the sequence of distributions of interest in this context is:
\begin{equation*}
\pi_l(u) = \frac{\kappa_l(u)}{Z_l} = \frac{e^{-\Phi[\cG(p^l(\cdot;u))]}}{\int_E e^{-\Phi[\cG(p^l(\cdot;u))]}du}\ , \quad l=1,\ldots, L.\ 
\end{equation*}
One is also interested in computing $Z_L$, for instance, to perform model selection or averaging.

Exact sampling of this sequence of posterior distributions is not possible in general,
and one must resort to an advanced method such as MCMC. 
But it is not obvious how one can leverage the MLMC approach for this application. 
Several strategies are suggested later on in the article.

\subsection{Partially Observed Diffusions}
\label{ssec:pod}

The following model is considered, as described in \cite{mlpf,mlpf1}.
Consider the 
partially-observed diffusion process:
\begin{eqnarray}
dU_t & = & a(U_t)dt + b(U_t)dW_t,
\label{eq:sde}
\end{eqnarray}
with $U_t\in\mathbb{R}^d$, $t\geq 0$, $U_0$ given 
$a:\mathbb{R}^d\rightarrow\mathbb{R}^d$ (denote the $j^{th}-$element as $a^j(U_t)$),
$b:\mathbb{R}^d\rightarrow\mathbb{R}^{d\times d}$ 
(denote the $j^{th},k^{th}-$element as $b^{j,k}(U_t)$)
and $\{W_t\}_{t\in[0,T]}$ a Brownian motion of $d-$dimensions. Some assumptions are made in \cite{mlpf,mlpf1} to ensure that the
diffusion has an appropriate solution; see \cite{mlpf,mlpf1} for details.

It will be assumed that the data are 
regularly spaced (i.e.,~in discrete time) observations 
$y_1,\dots,y_{n}$, with $y_k \in \bbR^m$.
It is assumed that conditional on $U_{k\delta}=u_{k\delta}$, 
for 
$1\ge \delta>0$,
$Y_k$ is independent of all other random variables and has density $G(u_{k\delta},y_k)$.
For simplicity of notation, let $\delta=1$ (which can always be done by rescaling time), 
so $U_k = U_{k\delta}$.
The joint probability density of the observations and the unobserved diffusion at the observation times 
is then
$$
\prod_{i=1}^n G(u_{i},y_{i})Q^\infty(u_{(i-1)},u_{i}), 
$$
where 
$Q^\infty(u_{(i-1)},u)$ 
is the transition density of the diffusion process as a function of $u$, 
i.e., the density of the solution
$U_1$ of Eq. \eqref{eq:sde} at time $1$
given initial condition 
$U_0=u_{(i-1)}$.

In this problem, one wants to \emph{sequentially} approximate a probability on a fixed space.
For $k\in\{1,\dots,n\}$, the objective is to approximate the filter
$$
\pi_{{\infty}}(u_{k}|y_{1:k}) = \pi^k_{{\infty}}(u_{k}) = \frac{\int_{\mathbb{R}^{(k-1)d}} \prod_{i=1}^k G(u_{i},y_{i})Q^\infty(u_{(i-1)},u_{i}) du_{1:k-1}}{\int_{\mathbb{R}^{kd}} \prod_{i=1}^k G(u_{i},y_{i})Q^\infty(u_{(i-1)},u_{i}) du_{1:k}} \, ,
$$
with $u_{1:k}=(u_1,\dots, u_k)$ and $y_{1:k}=(y_1,\dots,y_k)$.
The shorthand notation $\pi^k(\cdot) = \pi(\cdot|y_{1:k})$ is used above and in what follows.
Note that we will use $\pi_{{\infty}}$
as the notation for measure and density, with the use clear from the context.
It is also of interest, to estimate the normalizing constant,
or marginal likelihood 
$$
Z_{{\infty}}(y_{1:k}) = Z^k_{{\infty}} \int_{\mathbb{R}^{kd}} \prod_{i=1}^k G(u_{i},y_{i})Q^\infty(u_{(i-1)},u_{i}) du_{1:k} \, .
$$
Note that the filtering problem has many applications in engineering, statistics, finance, and physics (e.g.,~\cite{Cappe_2005,crisan_book, doucet_def} and the references therein)

\subsubsection{Approximation}\label{sec:euler_filter}

There are several issues associated to the approximation of the filter
and marginal likelihood, sequentially in time. Even if one knows
$Q^\infty$ pointwise, up-to a non-negative unbiased estimator, 
and/or can sample exactly from the associated law, 
advanced computational
methods, such as particle filters (e.g.,~\cite{doucet_johan,fearn})  -- an exchangeable term for SMC when used in a filtering context -- are often adopted in order to estimate the filter.
  In the setting considered in this paper, it is assumed that one cannot
\begin{itemize}
\item{evaluate $Q^\infty$ pointwise, up-to a non-negative unbiased estimator ;}
\item{sample from the associated distribution of $Q^\infty$.}
\end{itemize}
$Q^\infty$ and its distribution must be approximated by some discrete time-stepping method \cite{KlPl92} (for time-step $h_l=2^{-l}$).

For simplicity and illustration, Euler's method \cite{KlPl92} will be considered.
One has
\begin{eqnarray}
\label{eq:euler1step}
U_k^{l}(m+1) & = & U_k^{l}(m) + {h_l} a(U_k^{l}(m)) + \sqrt{h_l} b(U_k^{l}(m)) \xi_{k}(m) \, , \\
\xi_{k}(m) & \stackrel{\textrm{i.i.d.}}{\sim} & \mathcal{N}_d(0,I_d) \, ,
\nonumber\end{eqnarray}
for $m=0,\dots, k_l-1$, where 
$k_l=2^l$ and $\mathcal{N}_d(0,I_d)$ is the $d-$dimensional normal distribution with mean zero and covariance the identity (when $d=1$ we omit the subscript). 
Here $U_k^{l}(k_l)=U^l_k$, $U^l_k(0)=U^{l}_{k-1}=U_{k-1}^{l}(k_l)$.
The numerical scheme gives rise to its own transition density between observation times 
$Q^l(u^{l}_{k-1},u^l_k)$.

Therefore, one wants to approximate
for $k\in\{1,\dots,n\}$  the filter
$$
\pi_{{L}}(u_{k}|y_{1:k}) = \frac{\int_{\mathbb{R}^{(k-1)d}} \prod_{i=1}^k G(u_{i},y_{i})Q^L(u_{(i-1)},u_{i}) du_{1:k-1}}{\int_{\mathbb{R}^{kd}} \prod_{i=1}^k G(u_{i},y_{i})Q^L(u_{(i-1)},u_{i}) du_{1:k}} \, ,
$$
and  marginal likelihood 
$$
Z_L(y_{1:k}) = \int_{\mathbb{R}^{kd}} \prod_{i=1}^k G(u_{i},y_{i})Q^L(u_{(i-1)},u_{i}) du_{1:k}.
$$
First, we consider how this task can be performed using SMC 
and how that in turn can be extended to the MLMC context.

\subsubsection{Parameter Estimation}\label{sec:par_est}

Suppose that there is a static parameter $\theta\in\Theta\subseteq\mathbb{R}^{d_{\theta}}$ 
in the model, so 
$$
dU_t  =  a_{\theta}(U_t)dt + b_{\theta}(U_t)dW_t \, ,
$$
and $G_{\theta}$ is the likelihood function above. 
If one assumes a prior $\pi_{\theta}$ on $\theta$, 
then one might be interested in, for $k$ \emph{fixed}:
$$
\pi_{{\infty}}(d\theta|y_{1:k}) = 
\pi_{\theta}(d\theta)\frac{\int_{\mathbb{R}^{kd}} \prod_{i=1}^k G_{\theta}(u_{i},y_{i})Q_{\theta}^\infty(u_{(i-1)},u_{i}) du_{1:k}}{\int_{\mathbb{R}^{kd}\times\Theta} \prod_{i=1}^k G_{\theta}(u_{i},y_{i})Q_{\theta}^\infty(u_{(i-1)},u_{i}) \pi_{\theta}(d\theta)du_{1:k}}
$$
and the associated discretization. 
We consider how the latter task is possible using MCMC 
for a given level and how that is extended for the MLMC case.

\section{Some Computational Methods}\label{sec:comp_meth}

The following section gives a basic introduction to some computational methods that can be used for
the examples of the previous section. The review is quite basic, in the sense that there are numerous extensions
in the literature, but, we try to provide the basic ideas, with pointers to the literature, where relevant.
We first consider basic MCMC in Section \ref{sec:mcmc_sec}. Then, in Section \ref{sec:smc_sec}, SMC
is discussed in context of particle filtering. Here, a certain SMC algorithm, SMC samplers (e.g.,~\cite{delm:06})
is approached, which uses MCMC algorithms within it. In Section \ref{sec:pmcmc_sec}, we discuss particle MCMC \cite{andrieu}, which combines SMC within
MCMC proposals. Finally, in Section \ref{ssec:enkf}, we discuss ensemble Kalman filter approaches.

\subsection{Markov chain Monte Carlo}\label{sec:mcmc_sec}

We consider a target probability $\pi_L$ on measurable space $(\mathsf{E},\mathcal{E})=(\mathbb{R}^d,\mathcal{B}(\mathbb{R}^d))$, 
with $\mathcal{B}(\mathbb{R}^d)$ the Borel sets on $\mathbb{R}^d$; extensions to other spaces is simple, but it is omitted for brevity.
We write the target density of $\pi_L$ w.r.t.~Lebesgue measure as $\pi_L$ also, 
following standard practice.

The idea of MCMC is to build an ergodic Markov kernel of invariant measure $\pi_L$ or, at least, that $\pi_L$ is a marginal of the invariant
measure of the Markov chain -- we concentrate on the former case. That is, samples of the Markov chain $U^1,\dots,U^N$ have the property that for $\varphi:\mathsf{E}\rightarrow\mathbb{R}$,
$\varphi$ $\pi-$integrable,
one has that estimates of the form
$$
\frac{1}{N}\sum_{i=1}^N \varphi(u^i)
$$
will converge almost surely as $N\rightarrow\infty$ to $\mathbb{E}_{\pi_L}[\varphi(U)]$.

There are many ways to produce the Markov chain. Here we will only describe the standard random walk Metropolis-Hastings algorithm. 
At the end of the description, references
for more recent work are given. 
Suppose at time $i$ of the algorithm, $u^i$ is the current state of the Markov chain.
A new candidate state $U'|u^i$ is proposed according to
$$
U' = u^i + Z,
$$
where $Z\sim\mathcal{N}_d(0,\Sigma)$ independently of all other random variables. 
The proposed value is accepted ($u^{i+1}=u^i$) with probability:
$$
\min\Big\{1,\frac{\pi_L(u')}{\pi_L(u^i)}\Big\}
$$
otherwise it is rejected and $u^{i+1}=u^i$. 
The scaling of the proposal, i.e.,~the 
proposal covariance $\Sigma$,
is often chosen so that the acceptance rate is about 0.234 \cite{roberts1},
although there are adaptive methods for doing this; see~\cite{andrieu1,haario}.

The algorithm mentioned here is the most simple approach. The ideas can be extended to alternative proposals, Langevin \cite{roberts2}, Hamiltonian Monte Carlo (\cite{duane}, see also
\cite{neal}), pre-conditioned Crank
Nicholson (e.g.,~\cite{inverse_prob} and the references therein). The algorithm can also be used in infinite dimensions (e.g.,~Hilbert spaces~\cite{inverse_prob}) and each dimension need not be updated simultaneously - 
for example, one can use Gibbs and Metropolis-within-Gibbs approaches; 
see \cite{robert,roberts} for some coverage. 
There are also population-based methods (e.g.,~\cite{pop_based} 
and the references therein) and non-reversible approaches 
(e.g.,~\cite{ottobre1, ottobre2, non_rev}
and the references therein). 
Even this list is not complete: the literature is so vast, that one would require a book length introduction, which is well
out of the scope of this work - the reader is directed to the previous papers and the references therein for more information. There is also a well established convergence
theory; see \cite{meyn,roberts} for information. We remark also that the cost of such MCMC algorithms can be quite reasonable if $d$ is large, often of polynomial order
(e.g.,~\cite{roberts1}) and can be dimension free when there is a well-defined limit as $d$ grows
\cite{beskospcn, inverse_prob}. Note that finally, it is not simple to use `standard' MCMC to estimate normalizing
constants.

\subsection{Sequential Monte Carlo}\label{sec:smc_sec}

Here we will consider the standard SMC algorithm for filtering, 
also called the particle filter in the literature.
To assist our discussion, we focus on the 
filtering density from Section \ref{sec:euler_filter}
\begin{equation}\label{eq:approxfilt}
\pi^k_{{L}}(u_{k})= \pi_{{L}}(u_{k}|y_{1:k}) \propto  \int_{\bbR^{(k-1)d}}
\prod_{i=1}^{k} G(u_{i},y_{i})Q^L(u_{(i-1)},u_{i}) du_{1:k-1}\, . 
\end{equation}
Here there are no static parameters to be estimated. 
To facilitate the discussion, we will suppose that 
$Q^L$ can be evaluated pointwise, although of course, 
it is generally not the case in our application.
Moreover, it will be assumed that $G(u_{i},y_{i})Q^L(u_{(i-1)},u_{i}) >0$ 
for each $i\geq 1$, $u_{i-1},u_i$.

We suppose we have access to a collection of proposals 
$q_1(u_1),q_2(u_1,u_2), q_3(u_2,u_3),\dots$, where 
$q_j(u_{j-1},u_j)$ is a  
positive probability density in $u_j$, for each value of $u_{j-1}$. 
The particle filter is then as follows:
\begin{itemize}
\item{\textbf{Initialize}. Set $k=1$, for $i\in\{1,\dots,N\}$ sample $u_1^i$ from $q_1$ and evaluate the weight
$$
w_1^i = \Big(\frac{G(u_{1}^i,y_{1})Q^L(u_{0},u_{1}^i) }{q_1(u_1^i)}\Big)\Big(\sum_{j=1}^N \frac{G(u_{1}^j,y_{1})Q^L(u_{0},u_{1}^j) }{q_1(u_1^j)}\Big)^{-1}
$$
}
\item{\textbf{Iterate}: Set $k=k+1$,
\begin{itemize}
\item{Resample 
$(\hat{u}_{k-1}^1,\dots,\hat{u}_{k-1}^N)$ according to the weights $(w_{k-1}^1,\dots,w_{k-1}^N)$.}
\item{Sample $u_k^i|\hat{u}_{k-1}^i$ from $q_k$, for $i\in\{1,\dots,N\}$, 
and evaluate the weight
$$
w_k^i = \Big(\frac{G(u_{k}^i,y_{k})Q^L(\hat{u}_{k-1}^i,u_{k}^i) }{q_k(\hat{u}_{k-1}^i,u_k^i)}\Big)\Big(\sum_{j=1}^N 
\frac{G(u_{k}^j,y_{k})Q^L(\hat{u}_{k-1}^j,u_{k}^j) }{q_k(\hat{u}_{k-1}^j,u_k^j)}\Big)^{-1}.
$$
}
\end{itemize}
}
\end{itemize}

The resampling step can be performed using a variety of schemes, 
such as systematic, multinomial residual etc; the reader is referred to 
\cite{doucet_johan} for more details.
For $\varphi:\mathbb{R}^d\rightarrow\mathbb{R}$, 
and $\varphi$ $\pi^k_{{L}}-$integrable, one has the consistent estimate:
\begin{equation}\label{eq:smc_est}
\sum_{i=1}^N w_k^i\varphi(u_k^i) \approx  \bbE_{\pi^k_L}[\varphi(U_k)].
\end{equation}
In addition, the marginal likelihood is unbiasedly \cite{delm:04} estimated by
\begin{equation}\label{eq:nc_est}
\widehat{Z}_L^k := \prod_{i=1}^k \Big(\frac{1}{N}\sum_{j=1}^N \frac{G(u_{i}^j,y_{i})Q^L(\hat{u}_{i-1}^j,u_{i}^j) }{q_i(\hat{u}_{i-1}^j,u_i^j)}\Big) \approx Z_L^k,
\end{equation}
with the abuse of notation that $\hat{u}_0^i=u_0$. In principle, for $\varphi:\mathbb{R}^{kd}\rightarrow\mathbb{R}$, and $\varphi$ $\pi^k_{{L}}-$integrable, one could also try to estimate
$\mathbb{E}_{\pi^k_{{L}}}[\varphi(U_{1:k})]$ but this does not work well in practice due to the well-known path degeneracy problem; see \cite{doucet_johan,kantas}.

The algorithm given here is one of the most basic and many modifications can enhance the performance of this algorithm; see \cite{delm:04,delmoral1,doucet_johan,kantas} for some ideas.
The theoretical validity of the method has been established in many works; see e.g., \cite{cl-2013,chopin1,delm:04,delmoral1,douc}.
The algorithm performs very well w.r.t.~the time parameter $k$. Indeed $L_p-$errors for estimates such as \eqref{eq:smc_est} are $\mathcal{O}(N^{-1/2})$ where the constant
is independent of time and the relative variance of \eqref{eq:nc_est} is $\mathcal{O}(k/N)$ (if $N>Ck$ for $C$ some constant independent of $k$); see \cite{delmoral1} and the references therein. 
One of the main issues with particle filters/SMC methods in this context is that they do not perform well in high dimensions (i.e.,~for large $d$) often having an exponential
cost in $d$ \cite{snyder}. 
Note however, that if there is a well-defined limit as $d$ grows, 
SMC methods can be designed to perform quite well in practice 
on a finite time horizon (see e.g.,~\cite{kantas1} and the next section).
There have been some methods developed for high-dimensional filtering (e.g.,~\cite{beskos_hd,lind,rebs}), however, they are only useful for a small class of models.

\subsubsection{Sequential Monte Carlo Samplers}\label{sec:smc_samp_sec}

Consider a sequence of distributions $\pi_1,\dots,\pi_L$ on a common measurable space. 
In addition to this suppose we have Markov kernels $M_2,\dots,M_L$ of invariant measures
$\pi_2,\dots,\pi_L$. 
This is possible if the densities are known up-to a constant 
(and potentially also a non-negative unbiased estimator - 
although this is not considered at the moment), simply by using using MCMC. 
The SMC sampler algorithm (e.g.,~\cite{delm:06})
can be used to approximate expectations w.r.t.~$\pi_1,\dots, \pi_L$, 
as well as to estimate ratios of normalizing constants. 
The un-normalized densities (assumed to exist w.r.t.~a common dominating measure) of 
$\pi_1,\dots, \pi_L$ are written
$\kappa_1,\dots,\kappa_L$. 
To ease the notational burden, we suppose one can sample from $\pi_1$, but this is not necessary.

The algorithm is as follows:
\begin{itemize}
\item{\textbf{Initialize}. Set $l=1$, for $i\in\{1,\dots,N\}$ sample $u_1^i$ from $\pi_1$.}
\item{\textbf{Iterate}: Set $l=l+1$. If $l=L+1$ stop.
\begin{itemize}
\item{Resample
$(\hat{u}_{l-1}^1,\dots,\hat{u}_{l-1}^N)$ using the weights $(w_{l}^1,\dots,w_{l}^N)$
where, for $i\in\{1,\dots,N\}$, 
$$
w_l^i = \Big(\frac{\kappa_l(u_{l-1}^i)}{\kappa_{l-1}(u_{l-1}^i)}\Big)\Big(\sum_{j=1}^N \frac{\kappa_l(u_{l-1}^j)}{\kappa_{l-1}(u_{l-1}^j)}\Big)^{-1} .
$$
}
\item{Sample $u_l^i|\hat{u}_{l-1}^i$ from $M_l$ for $i\in\{1,\dots,N\}$.}
\end{itemize}
}
\end{itemize}
One can estimate expectations w.r.t.~$\pi_l$, 
for $\varphi:\mathsf{E}\rightarrow\mathbb{R}$, $\varphi$ $\pi_l-$integrable.
The consistent estimator
$$
\frac{1}{N}\sum_{i=1}^N \varphi(u_l^i) \approx \mathbb{E}_{\pi_l}[\varphi(U)]
$$
converges almost surely 
as $N\rightarrow\infty$. In addition, for any $l\geq 2$, we have the unbiased estimator
$$
\prod_{\ell=2}^l \Big(\frac{1}{N}
\sum_{i=1}^N\frac{\kappa_\ell(u_{\ell-1}^i)}{\kappa_{\ell-1}(u_{\ell-1}^i)}\Big) 
\approx Z_l/Z_1, 
$$
which converges almost surely 
as $N\rightarrow\infty$.

The basic algorithm goes back to at least \cite{jarzynski}. Several versions are found in \cite{chopin,neal:01}, with a unifying framework in \cite{delm:06} and a rediscovery in \cite{ching}. Subsequently, several refined and improved versions of the algorithm
have appeared \cite{chopin2,delmoralabc,heng,jasra_levy,shaefer}, including those which allow algorithmic parameters to be set adaptively, that is, without user specification.

Contrary to particle filters, when $\mathsf{E}=\mathbb{R}^d$, 
this method indeed performs quite well w.r.t.~the dimension $d$ with only polynomial cost in $d$; 
see \cite{beskos1,beskos_hd1}. Whilst the underlying theory
for this algorithm is very similar to particle filters and it is covered in \cite{delm:04,delmoral1}, 
there are some additional results in \cite{beskos_adap,schweizer,whiteley1}. In particular, \cite{beskos_adap} establish that 
when one updates parameters adaptively, 
such as in \cite{jasra_levy,shaefer}, then the algorithm is still theoretically correct. 
The method is very useful in the following scenaria:
(i) if one wishes to compute ratios of normalizing constants, 
(ii) 
the available MCMC kernels do not mix particularly well, and/or
(iii) the target is multimodal and the modes are separated by regions of very low probability.

\subsection{Particle Markov chain Monte Carlo}\label{sec:pmcmc_sec}

We now consider the scenario of Section \ref{sec:par_est}. 
In this context, the standard approach is to consider the extended target with density
$$
\pi_{{L}}^{k}(\theta,u_{1:k}) \propto
\pi_{\theta}(\theta) \prod_{i=1}^k G_{\theta}(u_{i},y_{i})Q_{\theta}^L(u_{(i-1)},u_{i}).
$$
Sampling this distribution is notoriously challenging. One recent method and it's extensions can be considered the gold standard as we
describe now. Note that the SMC algorithm (in Section \ref{sec:smc_sec}) used below uses the Euler discretized dynamics as the proposal.

The particle marginal Metropolis-Hastings (PMMH) algorithm of \cite{andrieu} proceeds as follows. 
\begin{itemize}
\item{\textbf{Initialize}. Set $i=0$ and sample $\theta^0$ from the prior. 
Given $\theta^0$ run the SMC algorithm in Section \ref{sec:smc_sec} and record the estimate of 
$\widehat{Z}_{L,\theta^0}^{k}$ 
from eq.~\eqref{eq:nc_est}.}
\item \textbf{Iterate}: 
\begin{itemize}
\item Set $i=i+1$ and propose $\theta'$ given $\theta^{i-1}$ from a proposal $r(\theta^{i-1},\cdot)$. \item Given $\theta'$ run the SMC algorithm in Section \ref{sec:smc_sec} 
and record the estimate $\widehat{Z}_{L,\theta'}^{k}$. 
\item Set $\theta^i=\theta'$ with probability
$$
\min\Big\{
1,
\frac{\widehat{Z}_{L,\theta'}^{k}
\pi_{\theta}(\theta')r(\theta',\theta^{i-1})}
{\widehat{Z}_{L,\theta^{i-1}}^{k}
\pi_{\theta}(\theta^{i-1})r(\theta^{i-1},\theta')}
\Big\}
$$
otherwise $\theta^i=\theta^{i-1}$.
\end{itemize}
\end{itemize}
The samples of this algorithm can be used to estimate expectations such as 
$\int_{\Theta}\varphi(\theta)\pi_{{L}}(\theta|y_{1:k})d\theta$ with
$$
\frac{1}{N}\sum_{i=1}^N \varphi(\theta^i) .
$$
Note that this is consistent as $N$ grows, in the sense that it recovers the true expectation with probability 1, under minimal conditions. The algorithm can
also be extended to allow estimation of the hidden states $u_{1:k}$ as well. 
There are many parameters of the algorithm, such as the number of samples of the SMC
algorithm, and tuning them has 
been discussed in \cite{andrieu,doucet}, for example.

The PMMH algorithm is the most basic in \cite{andrieu}. Several enhancements are in \cite{andrieu} and numerous algorithms that improve upon
this method can be found in~\cite{corr_pm,singh}.

\subsection{Ensemble Kalman filter}
\label{ssec:enkf}


The idea of the ensemble Kalman filter (EnKF) is to approximate the filtering 
distribution \eqref{eq:approxfilt} using an ensemble of particles and their 
sample covariance \cite{evensen1}.  As such they are sometimes also referred
to as sequential Monte Carlo methods, but we believe it is important to distinguish
them from the methods described in subsection \ref{sec:smc_sec}.
The observations are incorporated as though the process were linear and Gaussian,
hence requiring only the covariance approximation.
Hence, the method is consistent only in the case of a linear Gaussian model 
\cite{law2015data}.
However, it is robust even in high dimensions \cite{evensen} and can be tuned
to perform reasonably well in tracking and forecasting.  
It has therefore become very popular among practitioners.

It will be assumed here for simplicity
that the observation selection function is given by:
\begin{equation}\label{eq:gee}
G(u_i,y_i) \propto \exp(-\frac12 |\Gamma^{-\frac12}(H u_i - y_i) |^2 ) \, ,
\end{equation}
where $H$ is a linear operator.
The linearity assumption is without any loss of generality
since if $h$ is nonlinear, one can always extend the system to
$(u,v)^\top$, where $v=h(u)$.

The EnKF is executed in a variety of ways and only one will be considered here, 
the {\it perturbed observation} EnKF:

\[\mbox{Prediction}\;\;\;\;\left\{\begin{array}{lll}  
{u}_{j+1}^{(n)}&\sim Q^L(\widehat u_{j}^{(n)},\, \cdot \, ) \, , \;n=1,...,N,\vspace{4pt}\\
{m}_{j+1}&=\frac1N\sum_{n=1}^N{u}_{j+1}^{(n)} \, ,\vspace{4pt}\\
{C}_{j+1}&=\frac1{N-1}\sum_{n=1}^N({u}_{j+1}^{(n)}-
{m}_{j+1})({u}_{j+1}^{(n)}-{m}_{j+1})^T \, .
 \end{array}\right.
                                   \] 
\[\,\,\;\;\;\;\mbox{Analysis}\;\;\;\;\left\{\begin{array}{llll} S_{j+1}&=H{C}_{j+1}H^T+\Gamma, \vspace{4pt}\\ 
K_{j+1}&={C}_{j+1}H^TS_{j+1}^{-1},
\vspace{4pt}\\
\widehat u_{j+1}^{(n)}&=(I-K_{j+1}H){u}_{j+1}^{(n)}+K_{j+1}y_{j+1}^{(n)}, \;n=1,...,N, \vspace{4pt}\\
y_{j+1}^{(n)}&=y_{j+1}+\xi_{j+1}^{(n)}, \;n=1,...,N.                                    \end{array}\right.
                                     \] 
Here $\xi_j^{(n)}$ are i.i.d. \index{i.i.d.} draws from $N(0,\Gamma)$.
Perturbed observation refers to the fact that each particle sees an 
observation perturbed by an independent draw from $N(0,\Gamma)$. 
 This procedure ensures the Kalman Filter is obtained  
 in the limit of infinite ensemble in the linear Gaussian case \cite{law2015data}.  
Notice that the ensemble is not prescribed to be Gaussian, 
even though it is updated as though it were, so the limiting target
is some non-Gaussian $\widehat{\pi}_L^k$, which is in general 
not equal to the density defined by \eqref{eq:approxfilt} (see e.g., ~\cite{law2015deterministic}).  

\section{Approaches for MLMC Estimation}\label{sec:approach}

We now consider various ways in which the MLMC 
method can be used in these challenging situations, where 
it is non-trivial to construct couplings of the targets.

\subsection{Importance Sampling}\label{sec:by_is}

In this case, we investigate the ML 
identity where the sequence of targets $\pi_1,\dots,\pi_L$ are defined on a common measurable
space and are known up-to a normalizing constant; i.e.,~$\pi_l(u) = \kappa_l(u)/Z_l$ as in Section \ref{sec:bip} and \ref{sec:smc_samp_sec}.

In this scenario \cite{beskos} (see also \cite{beskos2,delm_sa,delm_mlnc,non_local}) investigate the simple modification
\begin{eqnarray}
\mathbb{E}_{\pi_L}[\varphi(U)] & = & \sum_{l=1}^L \{\mathbb{E}_{\pi_l}-\mathbb{E}_{\pi_{l-1}}\}[\varphi(U)]\nonumber\\
& = & 
\mathbb{E}_{\pi_1}[\varphi(U)] + \sum_{l=2}^{L}\mathbb{E}_{\pi_{l-1}}\Big[
\Big(\frac{\kappa_l(U)Z_{l-1}}{\kappa_{l-1}(U)Z_l} - 1\Big)\varphi(U)\Big]\label{eq:ml_is_id}.
\end{eqnarray}
The idea here is simple. If one does not know how to construct a coupling of the targets, then one replaces coupling by importance sampling.
The key point is that as the targets $\pi_l$ and $\pi_{l-1}$ are very closely related by construction, 
and therefore the change of measure formula above should
facilitate an importance sampling procedure that \emph{performs well}. 
Just as for `standard' MLMC (for instance as described in 
Section \ref{sec:ml_mlmc}) where the coupling has to be `good enough', 
the change of measure needs to be chosen appropriately to ensure that this approach can work well.
Recall from Section \ref{sec:smc_samp_sec} that 
SMC samplers can be designed to 
sequentially approximate 
$\pi_1,\dots,\pi_L$, and the ratios $Z_l/Z_{l-1}$.
Therefore the change of measure in \eqref{eq:ml_is_id} is very natural here.

The approach in \cite{beskos} is to simply run the algorithm of Section \ref{sec:smc_samp_sec}, except at step $l$ one resamples $N_{l+1}<N_{l}$ particles, where the 
schedule of numbers $N_{0:L-1}$ is chosen using a similar principle as for standard MLMC. 
The identity \eqref{eq:ml_is_id} can be approximated via:
\begin{equation}
\sum_{l=3}^{L}\Big\{\frac{
\sum_{i=1}^{N_{l-1}} \varphi(u_{l-1}^i)\frac{\kappa_l(u_{l-1}^i)}{\kappa_{l-1}(u_{l-1}^i)}}
{\sum_{i=1}^{N_{l-1}} \frac{\kappa_l(u_{l-1}^i)}{\kappa_{l-1}(u_{l-1}^i)}}
- 
\frac{1}{N_{l-1}}\sum_{i=1}^{N_1} \varphi(u_{l-1}^i)\Big\}
+\frac{\sum_{i=1}^{N_1} \varphi(u_1^i)\frac{\kappa_2(u_1^i)}{\kappa_1(u_1^i)}}
{\sum_{i=1}^{N_1} \frac{\kappa_2(u_1^i)}{\kappa_1(u_1^i)}}.\label{eq:is}
\end{equation}
Note that the algorithm need only be run up-to level $L-1$.
\cite{beskos} not only show that this is consistent, 
but also give a general MLMC theorem using the theory in 
\cite{delm:04} with some additional work and assumptions 
(which are relaxed in \cite{delm_mlnc}). In the context of the
example of Section \ref{sec:bip}, the authors show that the work to 
compute expectations relative to standard SMC samplers
(as in Section \ref{sec:smc_samp_sec}) is reduced to achieve a given MSE,
under the following assumptions.

\begin{ass}[MLSMC samplers] There is a $\varepsilon > 0$ such that
for all $l=1,\dots,L$, $u,v\in \mathsf{E}$, and $A \in \mathcal{E}$
\begin{itemize} 
\item $\varepsilon < \kappa_l(u) < \varepsilon^{-1}$ \, ;
\item $\varepsilon M_l(v,A) < M_l(u,A) < \varepsilon^{-1} M_l(v,A)$, 
where we recall from subsection \ref{sec:smc_samp_sec} 
that $M_l$ is the Markov kernel with invariant measure proportional to $\kappa_l$,
used to mutate the updated population of samples at step $l$.
\end{itemize}
\end{ass}

\begin{figure}\centering
  \includegraphics[scale=0.7]{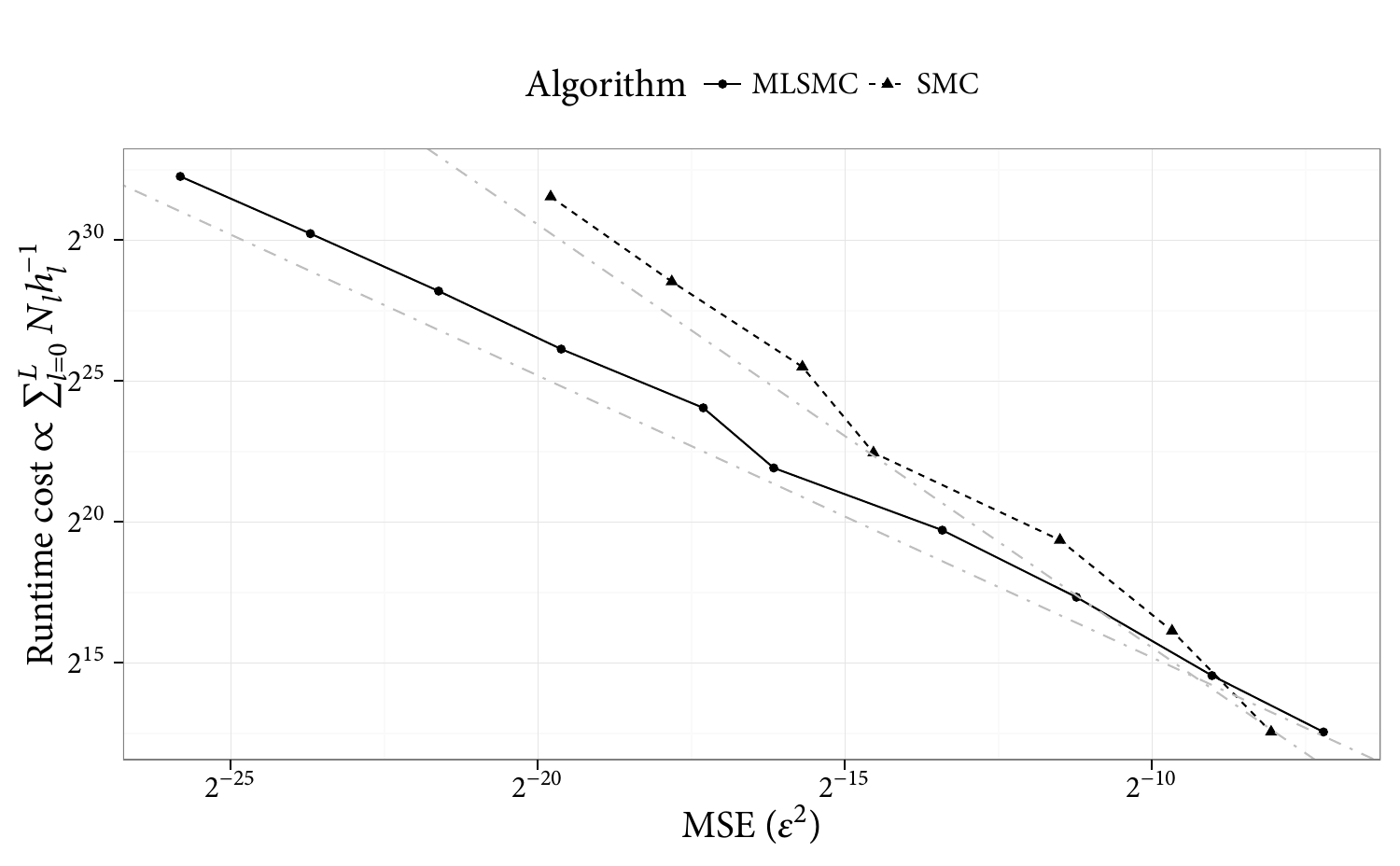}
  \caption{Computational cost against mean squared 
  error for MLSMC sampler in comparison to SMC sampler 
  for the Bayesian inverse problem from subsection \ref{sec:bip} 
  \cite{beskos}.}
  \label{fig:mlsmc}
\end{figure}

The main reason why this approach can work well, 
can be explained by terms that look like
$$
\frac{\kappa_l(U)Z_{l-1}}{\kappa_{l-1}(U)Z_l} - 1.
$$
In the context of the problem in Section \ref{sec:bip}, 
this term will tend to zero at a rate $h_l^{\beta}$, 
under suitable assumptions and in an appropriate norm, 
just as the variance terms in the coupling of Section \ref{sec:ml_mlmc}.
In more standard importance sampling language, 
the weight tends to one as the sequence of target distributions
gets more precise. 
This particular approach exploits this property, 
and the success of the method 
is dependent upon it. 
In particular, the key quantities for which one needs to obtain rates 
$\alpha$ and $\beta$ for are summarized in table \ref{tab:mlsmc}, 
and the convergence rate is illustrated in Figure \ref{fig:mlsmc}.

\begin{table}[h]
\begin{center}
  \begin{tabular}{ | c || c | c |}
    \hline
    Rate parameter & Relevant quantity  \\ 
    \hline\hline
    $\alpha$ & $(\bbE_{\pi_L} - \bbE_\pi)(\varphi)$  \\ 
    \hline
    $\beta$ & $\sup_{u\in \mathsf E} \left |\frac{\kappa_l(u) Z_{l-1}}{\kappa_{l-1}(u) Z_l} - 1 \right |$ \\ 
    \hline
  \end{tabular}
\end{center}
\caption{The key rates of convergence required for MLSMC samplers.}
\label{tab:mlsmc}
\end{table}

The method of \cite{beskos} has been extended to the computation of normalizing constants \cite{delm_mlnc} and has been applied to other examples,
such as non-local equations \cite{non_local} and 
approximate Bayesian computation \cite{ml_abc}. 
The method has also been extended
to the case that the accuracy of the approximation improves 
as the dimension of the target grows; see \cite{beskos2}.

The importance sampling idea has also been considered in other articles such as 
\cite{hoan:12, schicel}.
In \cite{hoan:12}, the change of measure appears in the context of MCMC 
and is not too dissimilar to the one presented in \eqref{eq:ml_is_id}, 
although it is with respect to the higher level in the couple, 
and multiple discretization indices are considered as well (see also \cite{mimc}).
In \cite{schicel}, the numerator and denominator of \eqref{eq:basic_exp}, 
arising from the form \eqref{eq:ratio}, are approximated independently
in terms of expectation w.r.t.~the prior, in a Bayesian set up,
and they find reasonably decent performance. 
This corresponds to a change of measure for Monte Carlo and Quasi Monte Carlo, 
although for MLMC it is slightly different, since MLMC is used separately for the
numerator and the denominator. 
In general such a procedure is not advisable unless the prior is very informative; 
one expects that the estimators for each of the numerator and the denominator 
will have very high variance.

\subsection{Approximate Coupling}\label{sec:by_ac}

The case of Section \ref{sec:par_est} will be considered here in order to illustrate this idea. 
In particular, there is a stochastic process that is partially observed. 
It is assumed that the dynamics of the associated discretized processes
can be easily coupled.  
Given this, an \emph{approximate} coupling is devised, which:
(i) can be sampled from (using MCMC/SMC), and (ii) has 
marginals which are similar to, but not exactly equal to, the pair $(\pi_l,\pi_{l-1})$. 
Then the difference $\{\mathbb{E}_{\pi_l}-\mathbb{E}_{\pi_{l-1}}\}[\varphi(U)]$
is replaced with an importance sampling formula (change of measure w.r.t.~the approximate coupling) and is approximated by sampling from the approximate coupling. 
The motivation for the idea will become clear as it is described in more detail.
The approach is considered in \cite{jasra} (see also \cite{jasra_mi}). 

We begin by considering:
$$
\pi_{{l}}^{k}(\theta,u_{1:k}) 
\propto
\pi_{\theta}(\theta) \prod_{i=1}^k G_{\theta}(u_{i},y_{i})Q_{\theta}^l(u_{(i-1)},u_{i})
$$
for two levels $l,l-1\geq 1$. We know that one can construct a good coupling of the two discretized kernels $Q_{\theta}^l,Q_{\theta}^{l-1}$ for any fixed
$\theta$ by sampling the finer Gaussian increments and concantenating them for the coarser discretization (e.g.,~\cite{giles} or as written in \cite{mlpf,mlpf1}). 
More precisly, given $\check{u}_{i-1}=(\underline{u}_{i-1},\overline{u}_{i-1})\in\mathbb{R}^{2d}$ and $\theta\in\Theta$, there is a Markov kernel $\check{Q}_{\theta}^{l,l-1}$
such that for any $A\in\mathcal{B}(\mathbb{R}^d)$ 
$$
\int_{A\times\mathbb{R}^d}\check{Q}_{\theta}^{l,l-1}(\check{u}_{i-1},\check{u}_{i})d\check{u}_{i} = \int_A Q^l_{\theta}(\overline{u}_{i-1},\overline{u}_i)d\overline{u}_i
$$
and 
$$
\int_{\mathbb{R}^d\times A}\check{Q}_{\theta}^{l,l-1}(\check{u}_{i-1},\check{u}_{i})d\check{u}_{i} = \int_A Q^{l-1}_{\theta}(\underline{u}_{i-1},\underline{u}_i)d\underline{u}_i.
$$
Note that under the coupling considered, the discretized processes are not independent.

We consider the joint probability on $\Theta\times\mathbb{R}^{2kd}$:
$$
\check{\pi}_{{l-1:l}}^{k}(\theta,\check{u}_{1:k}) \propto
\pi_{\theta}(\theta) \prod_{i=1}^k \check{G}_{\theta}(\check{u}_{i},y_{i})\check{Q}_{\theta}^{l,l-1}(\check{u}_{i-1},\check{u}_{i})
$$
for any non-negative function $\check{G}_{\theta}(\check{u}_{i},y_{i})$. 
Whilst this function can be `arbitrary', up-to some
constraints, we set it as 
\begin{equation}\label{eq:approxcoup}
\check{G}_{\theta}(\check{u}_{i},y_{i})=\max\{G_{\theta}(\overline{u}_{i},y_{i}),G_{\theta}(\underline{u}_{i},y_{i})\}.
\end{equation} 
This will be explained below. 
Let $\varphi:\Theta\times\mathbb{R}^{2kd}\rightarrow \mathbb{R}$ be 
$\pi_{l}^k$ and $\pi_{{l-1}}^k-$integrable.
Then we have (supressing the conditioning on $y_{1:k}$ in the expectations)
$$
\mathbb{E}_{\pi_{l}^k}[\varphi(\theta,U_{1:k})] - \mathbb{E}_{\pi_{{l-1}}^k}[\varphi(\theta,U_{1:k})] = 
$$
\begin{equation}\label{eq:approxis}
\frac{\mathbb{E}_{\check{\pi}_{{l-1:l}}^{k}}[\varphi(\theta,\overline{U}_{1:k})\overline H_{\theta}(\check{U}_{1:k})]}{\mathbb{E}_{\check{\pi}_{{l-1:l}}^{k}}[\overline{H}_{\theta}(\check{U}_{1:k})]} -
\frac{\mathbb{E}_{\check{\pi}_{{l-1:l}}^{k}}[\varphi(\theta,\underline{U}_{1:k})\underline H_{\theta}(\check{U}_{1:k})]}{\mathbb{E}_{\check{\pi}_{{l-1:l}}^{k}}[\underline{H}_{\theta}(\check{U}_{1:k})]} \, ,
\end{equation}
where
\begin{eqnarray*}
\overline{H}_{\theta}(\check{u}_{1:k}) & = & \prod_{i=1}^k \frac{ G_{\theta}(\overline{u}_{i},y_{i}) }{\check{G}_{\theta}(\check{u}_{i},y_{i})} \, , \\
\underline{H}_{\theta}(\check{u}_{1:k}) & = & \prod_{i=1}^k\frac{G_{\theta}(\underline{u}_{i},y_{i})}{\check{G}_{\theta}(\check{u}_{i},y_{i})} \, .
\end{eqnarray*}
The difference can then be approximating by sampling from $\check{\pi}_{{l-1:l}}^{k}$,~e.g.,~by using the PMMH \footnote{More precisely, one samples from a suitably extended measure, as described in \cite{andrieu}.} from Section \ref{sec:pmcmc_sec}.
This is done independently for each summand in the ML identity, with the first summand (the coarsest discretization) sampled by PMMH.

\begin{figure}
  \includegraphics[width=\linewidth]{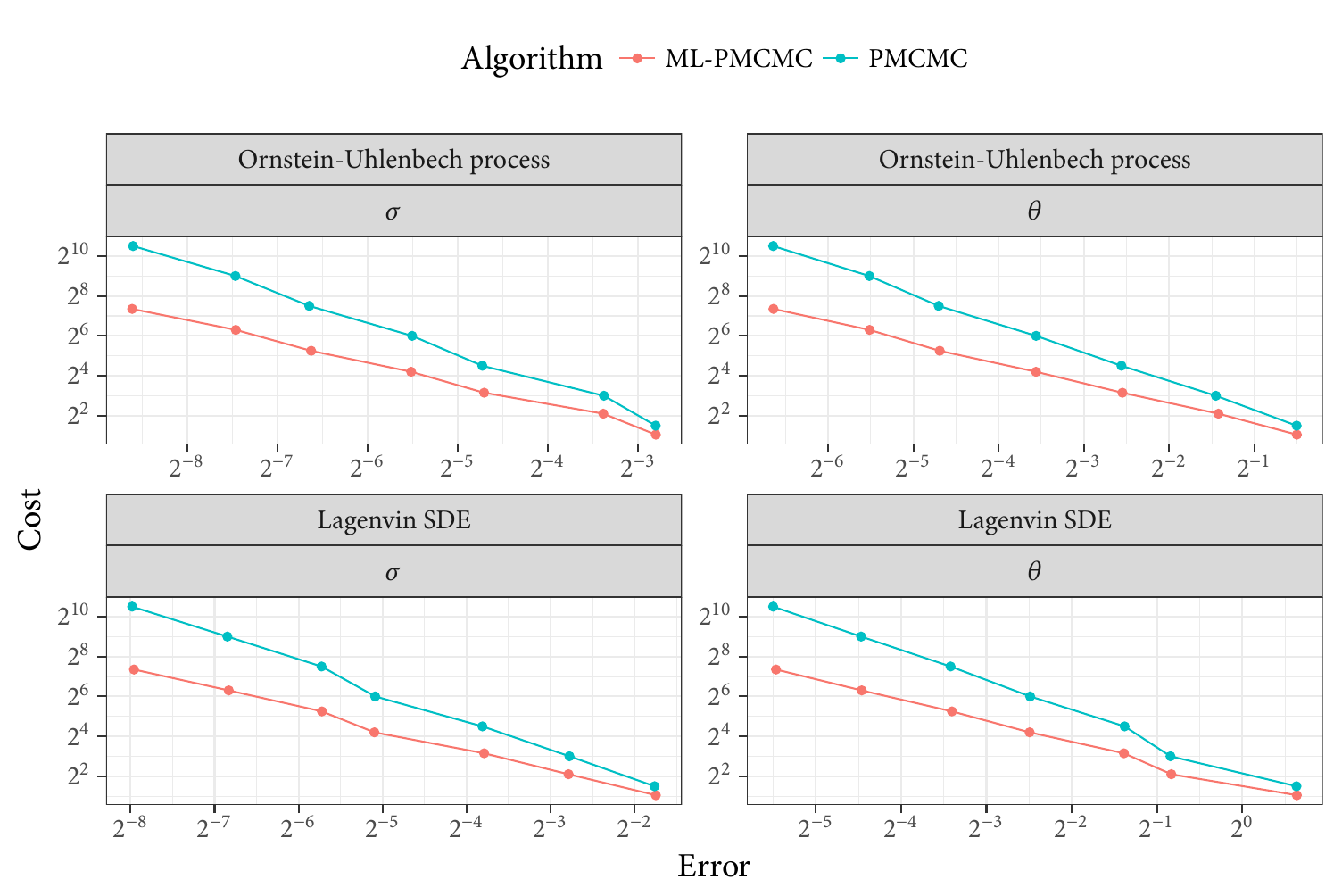}
  \caption{Cost vs. MSE for the inference of 2 parameters 
  for each of 2 SDE examples \cite{jasra}.}
  \label{fig:bpe}
\end{figure}

We now explain the idea in more detail. The basic idea is 
that one knows how to construct an exact coupling of the discretizations 
of the prior, i.e.,~the stochastic forward dynamics here, 
and it is natural to leverage this.
However, as noted previously, exact couplings of the posterior are not trivial to sample.
Instead, one aims to construct a joint probability that should have marginals which are close to,
but not exactly equal to, the correct ones.
As the coupling is not exact, one must correct for this fact and, as in Section \ref{sec:by_is}, 
use importance sampling. Just as argued in that section, the associated
weights of the importance sampling, that is, 
the terms $(\overline H_{\theta},\underline H_{\theta})$ 
should be well behaved in some sense. 
This can be ensured by choosing the function 
$\check{G}_{\theta}$ so 
that the variance of the weights 
w.r.t.~any probability measure will remain bounded 
uniformly in time. 
Hence the reason for its selection as \ref{eq:approxcoup}. 
\cite{jasra} are able to prove, under suitable assumptions on the model and PMMH kernel, 
that the computational effort to estimate a class of expectations is reduced versus a single PMMH algorithm on the finest level, for a given MSE sufficiently small.
The 
reduction in cost 
is a direct consequence of the prior coupling and well-behaved importance weights,
in connection with the ML identity. 
Note that the results of \cite{jasra} do not consider the 
dependence on the time parameter $k$, and that is something that
should be addressed.  In particular, the required assumptions in this context are
given below.
\begin{ass}[(PMMH using MLMC)] There is a $\varepsilon > 0$ 
and probability $\nu$ over $\Theta \times \bbR^{2kd}$, such that
for all $l=1,\dots,L$, $u\in \bbR^d$, $y\in \bbR^m$, $\theta \in \Theta$,
$A \in \sigma(\Theta)$, and any $\varphi : \Theta \times \bbR^{2kd}$ bounded 
and Lipschitz, and $w \in \mathsf{W}$ 
(the space of all auxiliary variables involved in the higher-dimensional chain),
the following hold
\begin{itemize} 
\item $\varepsilon < G_\theta(u,y) < \varepsilon^{-1}$ \, ;
\item $Q^l(u,v) > \varepsilon$ \, ;
\item The final Metropolis kernel $K$ on the extended space satisfies 
$ \int_\mathsf{W} \varphi(w')K(w,dw') \geq 
\varepsilon \int_{\Theta \times \bbR^{2kd}} \varphi(v) \nu(dv)$, 
\end{itemize}
\end{ass}
The quantities for which rates $\alpha$ and $\beta$ 
need to be obtained in this context are given in table \ref{tab:mlbpe}, 
and the results are illustrated in Figure \ref{fig:bpe} for 2 example SDE
of the type introduced in section \ref{sec:par_est}, 
each with 2 unknown parameters $\sigma$ and $\theta$.
\begin{table}[h]
\begin{center}
  \begin{tabular}{ | c || c | c |}
    \hline
    Rate parameter & Relevant quantity  \\ 
    \hline\hline
    $\alpha$ & $(\bbE_{\pi^k_L} - \bbE_{\pi^k})(\varphi)$  \\ 
    \hline
    $\beta$ & 
    $\int_{\Theta \times \bbR^{2kd}} 
    |\varphi(\theta,\overline{u}_{1:k}) - \varphi(\theta,\underline{u}_{1:k})|^2
    \check{\pi}_{{l-1:l}}^{k}(\theta,\check{u}_{1:k}) 
    d\theta d\check{u}_{1:k}$ \\ 
    \hline
  \end{tabular}
\end{center}
\caption{The key rates of convergence required for PMMH using MLMC.}
\label{tab:mlbpe}
\end{table}

We end this section by mentioning that the strategy described in this section 
is not the only one that could be adopted. 
For instance, the importance sampling approach of the previous section might also be considered.
Indeed that may be considered as an extreme example of approximate coupling in which 
$\check{G}_{\theta}(\check{u}_{i},y_{i})
= G_{\theta}(\underline{u}_{i},y_{i}) \delta_{\overline{u}_{i},\underline{u}_{i}}.$
There are some reasons why the approximate coupling method 
described in this section might be preferred, for the example considered here. 
Firstly, the terms $\kappa_l(u)/\kappa_{l-1}(u)$ are not available pointwise for this example; 
this could possibly be dealt with by random weight ideas (e.g.,~\cite{glass,rousset}), 
but it is still an issue. 
Secondly, there is a well-designed MCMC algorithm for the target (i.e.,~a PMMH kernel) 
and hence one would like to use this, 
as it is one of the gold standards for such models. 
If one elects to use an SMC sampler to approximate \eqref{eq:approxis},
then PMMH kernels can be used 
as described in \cite{chopin2}. 
The algorithm of \cite{chopin2} 
can also be used for dynamic (sequential) inference on the parameter, 
in an MLMC context.  
The same principle as described above can be generalized to MCMC,
and this has been done in the work \cite{ourmimcmc}.

\subsection{Coupling Algorithms}\label{sec:by_ca}

We now consider the case where one seeks to approximate 
the differences in the ML identity exactly. 
This is achieved by somehow trying to \emph{correlate} or couple stochastic algorithms,
rather than by constructing any joint coupling, either exact or approximate. 
We refer to this approach as coupling algorithms and it is explained further below.

\subsubsection{Coupling MCMC}

We begin by considering the method in \cite{scheichlmlmcmc2013}, 
which is a Markov chain approach. 
Consider two probability measures $(\pi_l,\pi_{l-1})$, where the support
of $\pi_{l-1}$ is $\mathsf{E}$ and the support of $\pi_l$ is 
$\mathsf{E}\times\mathsf{U}$. 
We focus on computing 
$\mathbb{E}_{\pi_l}[\varphi_l(U)] - \mathbb{E}_{\pi_{l-1}}[\varphi_{l-1}(U)]$ where $\varphi_l:\mathsf{E}\times\mathsf{U}\rightarrow\mathbb{R}$
and $\varphi_{l-1}:\mathsf{E}\rightarrow\mathbb{R}$ are $\pi_l$ and $\pi_{l-1}$ integrable.

Suppose we have a current state $(u_l,u_{l-1})\in\mathsf{E}\times\mathsf{U}\times\mathsf{E}$.
The approach consists first by sampling from $\pi_{l-1}$ exactly. Given this sample and the current $u_{l}\in\mathsf{E}$ a new
state is proposed from a proposal $r$ 
and is accepted or rejected according to the standard Metropolis-Hastings method.
It is clear that the samples are coupled and seemingly that they have the correct invariant distribution. \cite{scheichlmlmcmc2013} show that
this approach indeed can obtain the advantages of MLMC estimation for some examples. 
As noted in that article, supposing exact sampling from $\pi_{l-1}$ is
feasible, is not realistic and the authors propose a subsampling method to assist with this. See \cite{scheichlmlmcmc2013} for more details.

Before concluding this section, we mention the recent related works 
\cite{
jacob,coup_pf,spru}.
The work \cite{spru} (see also \cite{giles2}) considers coupled
Stochastic Gradient Langevin algorithms for some i.i.d.~models in Bayesian statistics. 
Note that there the levels are from an Euler discretization associated
to the algorithm, not the model per-se as described here. 

\subsubsection{Coupling Particle Filters}
\label{sssec:mlpf}

We consider the context of Section \ref{sec:euler_filter} 
of filtering a discretely and partially observed diffusion processes.
We describe an approach in \cite{mlpf} (see also \cite{reich,jacob,mlpf1,coup_pf}). 
For $l\geq 2$, $\varphi:\mathbb{R}^d\rightarrow\mathbb{R}$, 
$\pi^k_{{l}}$ and $\pi^k_{{l-1}}-$integrable
we consider approximating 
the difference $\mathbb{E}_{\pi^k_{{l}}}[\varphi(U_k)]-\mathbb{E}_{\pi^k_{{l-1}}}[\varphi(U_k)]$ sequentially in time.
Some of the notations of Section \ref{sec:by_ac} are also used.
The parameter $\theta$ is also dropped from the notations, as it is assumed 
to be fixed here.

The  multilevel particle filter (MLPF) is described as follows.
First, for $l=1$, run a particle filter for the coarsest discretization.
Now, run the 
following procedure independently for each $l\geq 2$.

{\bf For} $i=1,\dots, N_l$, draw $(\overline{U}^{l,i}_{1},\underline{U}_{1}^{l,i})\stackrel{\textrm{i.i.d.}}{\sim} \check{Q}^{l,l-1}((u_0,u_0),\cdot)$.

{\bf Initialize} $k=1$.  {\bf Do}

\begin{itemize}
\item[(i)] {\bf For} $i=1,\dots, N_l$, draw $(\overline{I}_{k}^{l,i},\underline{I}_{k}^{l,i})$ according to the coupled resampling procedure below. Set
$k = k+1$.
\item[(ii)] {\bf For} $i=1,\dots, N_l$, independently draw $(\overline{U}^{l,i}_{k},\underline{U}^{l,i}_{k})| (\overline{u}_{k-1}^{l,\overline{I}_{k}^{l,i}},
\underline{u}_{k-1}^{l,\underline{I}_{k}^{l,i}})\sim \check Q^{l,l-1}((\overline{u}_{k-1}^{l,\overline{I}_{k}^{l,i}},
\underline{u}_{k-1}^{l,\underline{I}_{k}^{l,i}}),~\cdot~)$.
\end{itemize}

The coupled resampling procedure for
the indices $(\overline{I}^{l,i}_{k}, \underline{I}^{l,i}_{k})$  is described below. 
First let \begin{equation}
\overline{w}_{k}^{l,i} = \frac{G(\overline{u}^{l,i}_{k},y_{k})}{\sum_{j=1}^{N_l} G(\overline{u}^{l,j}_{k},y_{k})} \qquad {\rm and} \qquad
\underline{w}_{k}^{l,i} = \frac{G(\underline{u}^{l,i}_{k},y_{k})}{\sum_{j=1}^{N_l} G(\underline{u}^{l,j}_{k},y_{k})} \, .
\label{eq:weights}
\end{equation}
Now
\begin{itemize}
\item[{\bf a}.] with probability  $\alpha_k^l = \sum_{i=1}^{N_l}\overline{w}_{k}^{l,i}\wedge \underline{w}_{k}^{l,i}$, 
draw $\overline{I}^{l,i}_{k}$ according to
$$
\bbP(\overline{I}^{l,i}_{k}=j) = \frac{1}{\alpha_k^l} (\overline{w}_{k}^{l,j}\wedge \underline{w}_{k}^{l,j}),
\qquad j\in\{1,\ldots,N_l\} \, ,
$$
and let $\underline{I}^{l,i}_{k}=\overline{I}^{l,i}_{k}$.
\item[{\bf b}.] otherwise, draw
$(\overline{I}^{l,i}_{k},\underline{I}^{l,i}_{k})$ independently according to the probabilities 
\[
\begin{split}
\bbP(\overline{I}^{l,i}_{k}=j) & =  [\overline{w}_{k}^{l,j}-
\overline{w}_{k}^{l,j}\wedge \underline{w}_{k}^{l,j}]/(\sum_{s=1}^{N_l}\overline{w}_{k}^{l,s}-\overline{w}_{k}^{l,s}\wedge \underline{w}_{k}^{l,s}) \, , \qquad j\in\{1,\ldots,N_l\} \, , \\ 
\bbP(\underline{I}^{l,i}_{k}=j) & = [\underline{w}_{k}^{l,j}-
\overline{w}_{k}^{l,j}\wedge \underline{w}_{k}^{l,j}]/(\sum_{s=1}^{N_l}\underline{w}_{k}^{l,s}-\overline{w}_{k}^{l,s}\wedge \underline{w}_{k}^{l,s}) \, , \qquad j\in\{1,\ldots,N_l\} \, .
\end{split}
\]
\end{itemize}

Note that by using 
the coupled kernel $\check Q^{l,l-1}$, one 
is sampling from the exact coupling of the discretized process, 
$(\overline{U}^{l,i}_{k},\underline{U}^{l,i}_{k})$.
Now one wants
to maintain 
as much dependence as possible in the resampling, 
since resampling is necessary in particle filters. 
The coupled resampling 
described above maximizes the probability (conditional on the history)
that the pair of samples remain coupled (see also \cite{chopin3}).

In the work \cite{mlpf}, it is shown that
$$
\sum_{i=1}^{N_l} \Big\{\varphi(\overline{u}_{k}^{l,i})\overline{w}_{k}^{l,i} - \varphi(\underline{u}_{k}^{l,i})\underline{w}_{k}^{l,i}\Big\}
$$
consistently approximates 
$\mathbb{E}_{\pi^k_{{l}}}[\varphi(U_k)]-\mathbb{E}_{\pi^k_{{l-1}}}[\varphi(U_k)]$. 
The MLPF estimator of $\bbE_{\pi^k_L}[\varphi(U_k)]$ is therefore given by 
$$
\sum_{i=1}^{N_1} w_k^{1,i} \varphi(u_k^{1,i}) 
+ \sum_{l=2}^L \sum_{i=1}^{N_l} \Big\{\varphi(\overline{u}_{k}^{l,i})\overline{w}_{k}^{l,i} - \varphi(\underline{u}_{k}^{l,i})\underline{w}_{k}^{l,i}\Big\}.
$$

In the case of Euler-Maruyama discretization, 
\cite{mlpf} it is shown that under suitable assumptions and for finite time
the standard choice of $L$ and $N_{1:L}$ as in \eqref{eq:biasL}
and \eqref{eq:nell} provides 
an MSE of $\mathcal{O}(\epsilon^2)$ 
for a cost of $\mathcal{O}(\epsilon^{-2.5})$.
For a particle filter the cost required 
is $\mathcal{O}(\epsilon^{-3})$. 
The theory is not limited to Euler discretizations, 
but the ultimate bound on the cost will
depend on the convergence rate of the numerical method.

\begin{figure}
 \includegraphics[width=\columnwidth]{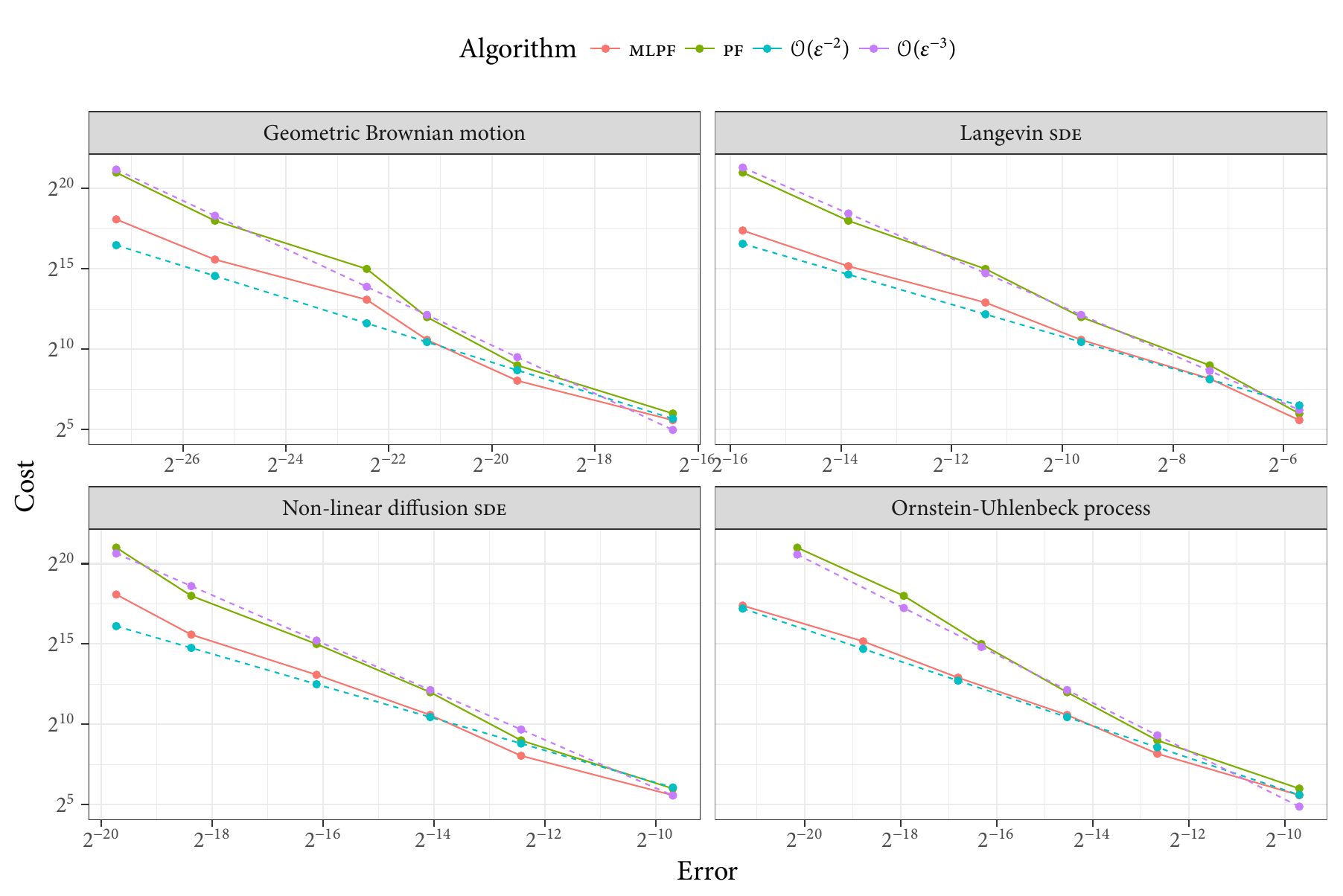}
  \caption{Cost rates as a function of MSE using MLPF in comparison 
  to bootstrap particle filter for  
  various SDE examples of the form \eqref{eq:sde} \cite{mlpf}.}
  \label{fig:mlpf}
\end{figure}

Sufficient assumptions in this case are given by 
\begin{ass}[MLPF] There is a $\varepsilon > 0$ such that
for all $l=1,\dots,L$, $u,v\in \bbR^d$, and $y\in \bbR^m$, 
the following hold
\begin{itemize} 
\item $\varepsilon < G_\theta(u,y) < \varepsilon^{-1}$ \, ;
\item $Q^l(u,v) > \varepsilon$.
\end{itemize}
\end{ass}
The quantities for which rates $\alpha$ and $\beta$ 
need to be obtained in this context are given in table \ref{tab:mlbpe},
and the complexity results are illustrated in Figure \ref{fig:mlpf} for
some example SDEs.
\begin{table}[h]
\begin{center}
  \begin{tabular}{ | c || c | c |}
    \hline
    Rate parameter & Relevant quantity  \\ 
    \hline\hline
    $\alpha$ & $(\bbE_{\pi^k_L} - \bbE_{\pi^k})(\varphi)$  \\ 
    \hline
    $\beta$ & 
    $\left(\int_{\bbR^{2d}} 
    |\varphi(\overline{u}_{k}) - \varphi(\underline{u}_{k})|^2
    \check{\pi}_{{l-1:l}}^{k}(\check{u}_{k}) 
    d\check{u}_{k}\right)^2$ \\ 
    \hline
  \end{tabular}
\end{center}
\caption{The key rates of convergence required for MLPF.}
\label{tab:mlpf}
\end{table}

In \cite{mlpf1} the approach and results are extended to the case of marginal likelihood estimation. We note that one drawback of the mathematical results
in \cite{mlpf,mlpf1} is that they do not consider the time-parameter. 
Note also that the method does not the preserve
the standard squared strong convergence of the forward 
numerical method which defines $\check Q^{l,l-1}$.  
A power of $1/2$ is lost in the rate of convergence 
as a result of the coupled resampling (see (table \ref{tab:mlpf}).
In \cite{jacob,coup_pf} the coupled resampling method is improved 
by using optimal transportation techniques \cite{villani}. 
Also \cite{reich} (see also \cite{greg}) obtain empirical results which indicate that more favorable convergence rates may be preserved in certain cases 
by replacing the resampling step with a deterministic linear transformation
of the current population, derived from the optimal transportation. 
However, in \cite{reich,greg,jacob,coup_pf} 
there are no mathematical results which
support the encouraging empirical results; 
one expects that this is mainly a technical challenge.

\subsubsection{Coupling the EnKF}

As discussed in subsection \ref{ssec:enkf}, the EnKF targets a different 
distribution than the filtering distribution in general, which has been denoted
$\widehat{\pi}^k_L$.
In between updates, this algorithm proceeds similarly to the MLPF of the previous 
section, propagating pairs of ensembles for each $l\geq 2$.
The fundamental difference in the update results in approximations
of increments $\mathbb{E}_{\widehat\pi^k_{{l}}}[\varphi(U_k)]-\mathbb{E}_{\widehat\pi^k_{{l-1}}}[\varphi(U_k)]$.
Therefore, the MSE will ultimately depend upon the difference 
$\mathbb{E}_{\widehat\pi^k_{{L}}}[\varphi(U_k)] - \mathbb{E}_{\pi^k_{{L}}}[\varphi(U_k)]$, 
which includes a Gaussian bias in addition to the discretization bias.
In order to preserve the coupling of this algorithm after the update, the sample 
covariance is approximated using the entire multilevel ensemble in \cite{mlenkf}, 
as follows.  Recall the functions $G(u_i,y_i)$ 
are assumed to take the form given in \eqref{eq:gee}.

{\bf For} $l=2,\dots,L$, and $i=1,\dots, N_l$,  
draw $(\overline{U}^{l,i}_{1},\underline{U}_{1}^{l,i})\stackrel{\textrm{i.i.d.}}{\sim} \check{Q}^{l,l-1}((u_0,u_0),\cdot)$.
And draw $U^{1,i}_1 \sim Q^1(u_0, \cdot )$.

{\bf Initialize} $k=1$.  {\bf Do}

\begin{itemize}
\item[(i)] Compute the MLMC covariance estimator \cite{chernov} :
\[
\begin{split}
{C}_k^{\rm ML} = & \frac1{N_1} \sum_{i=1}^{N_1} 
{u}^{1,i}_{k}({u}^{1,i}_{k})^T - \left(\frac1{N_1} \sum_{i=1}^{N_1} { u}^{1,i}_{k}\right)
\left(\frac1{N_1} \sum_{i=1}^{N_1} { u}^{1,i}_{k}\right)^T \\
+ & \sum_{l=2}^L \Biggr [
\frac1{N_l} \sum_{i=1}^{N_l} 
\left( {\overline u}^{l,i}_{k}({\overline u}^{l,i}_{k})^T - {\underline u}^{l,i}_{k}({\underline u}^{l,i}_{k})^T \right )
- \\
& \left(\frac1{N_l} \sum_{i=1}^{N_l} {\overline u}^{l,i}_{k}\right)\left(\frac1{N_l} \sum_{i=1}^{N_l} {\overline u}^{l,i}_{k}\right)^T
+ \left(\frac1{N_l} \sum_{i=1}^{N_l} {\underline u}^{l,i}_{k}\right)
\left(\frac1{N_l} \sum_{i=1}^{N_l} {\underline u}^{l,i}_{k}\right)^T \Biggr ] \, .
\end{split}
\]
\vspace{.1pt}
\item[(ii)] Compute $K_k^{\rm ML} = {C}_{k}^{\rm ML}H^T(H{C}_{+,k}^{\rm ML}H^T+\Gamma)^{-1}$, where $C_{+,k}^{\rm ML}$ the positive semi-definite modification of 
${C}_{k}^{\rm ML}$.
\item[(iii)] {\bf For} $l=2,\dots,L$, and $i=1,\dots, N_l$, 
independently draw $Y_{k}^{l,i} \sim N(y_{k}, \Gamma)$, 
and compute
$$
\widehat{\overline{u}}_{k}^{i}=(I-K_{k}^{\rm ML}H){\overline{u}}_{k}^{l,i}+K_{k}^{\rm ML}y_{k}^{l,i} \, ,
$$
and similarly for $\widehat{\underline{u}}_{k}^{l,i}$ and $\widehat{u}_{k}^{1,i}$.
Set $k = k+1$.
\item[(iv)] {\bf For} $l=2,\dots,L$, and $i=1,\dots, N_l$, independently draw 
$(\overline{U}^{l,i}_{k},\underline{U}_{k}^{l,i})\stackrel{\textrm{i.i.d.}}{\sim} 
\check{Q}^{l,l-1}((\widehat{\overline{u}}_{k-1}^{i}, \widehat{\underline{u}}_{k-1}^{i}),\cdot)$.
And draw $U^{1,i}_k \sim Q^1(\widehat{u}_{k-1}^{1,i}, \, \cdot \, )$.
\end{itemize}

\begin{figure}[htbp]
  \centering
  \includegraphics[width=1\textwidth]{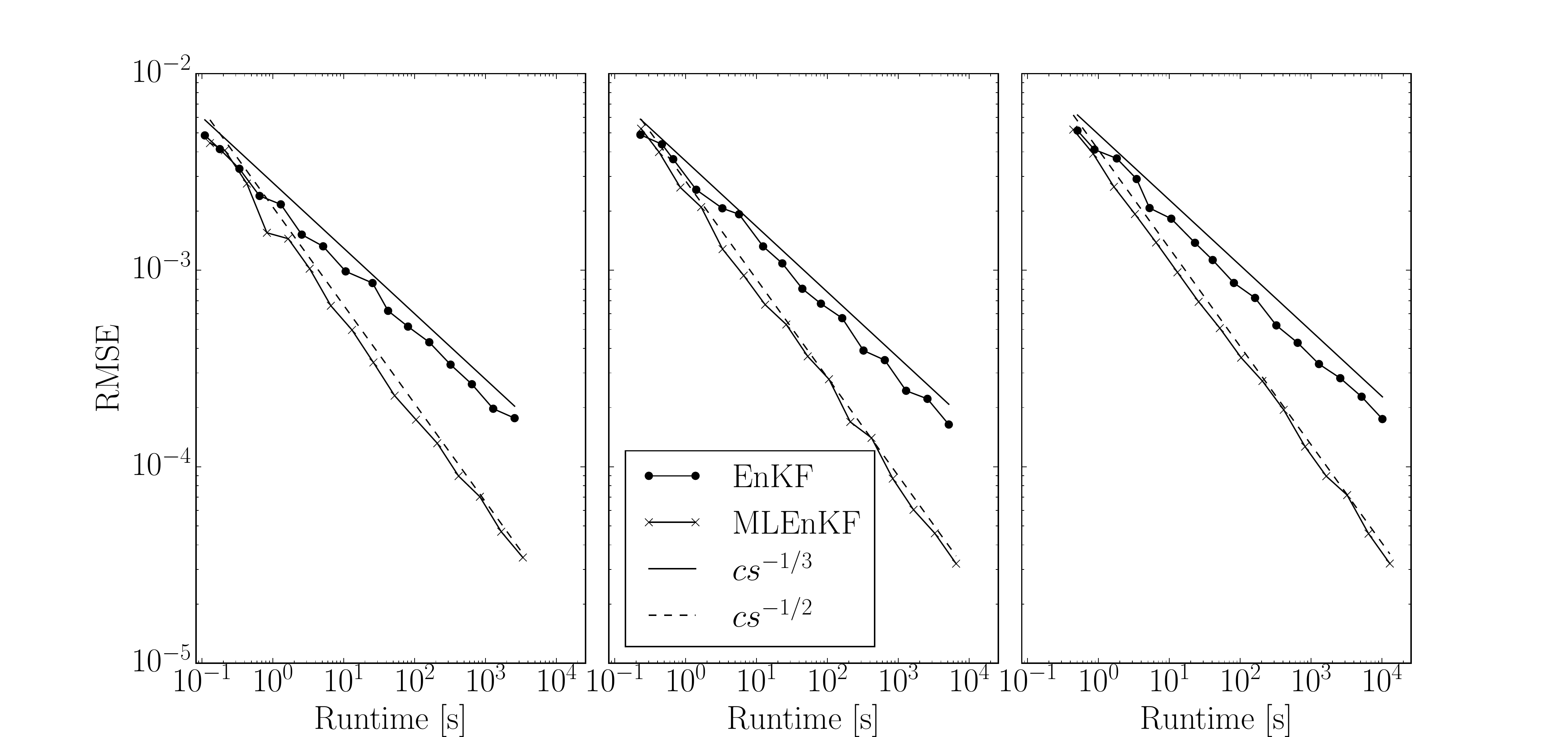}
   \includegraphics[width=1\textwidth]{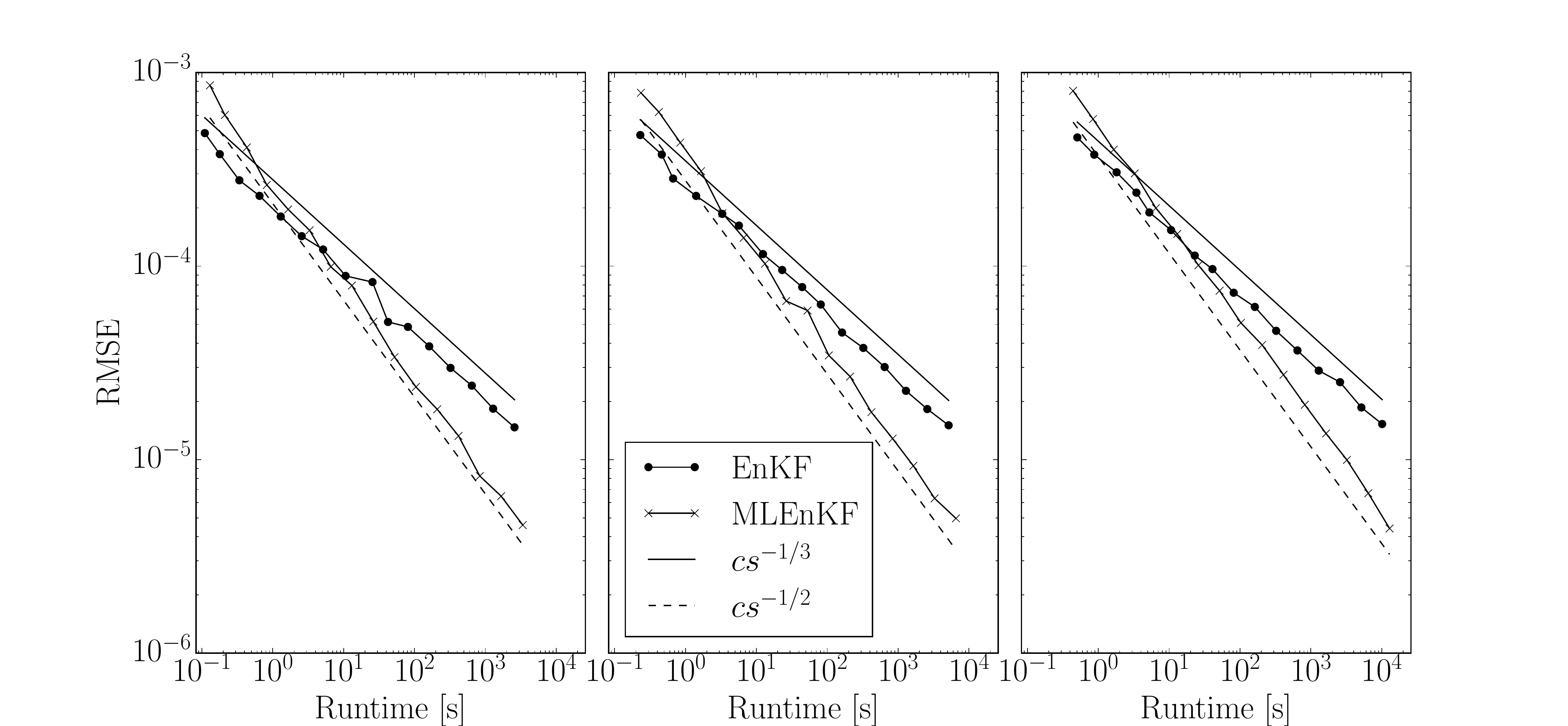}
  \caption{Comparison of the accuracy vs.~computational cost when using
  the EnKF and MLEnKF methods on a linear Gaussian filtering problem 
  of the form given in \eqref{eq:sde}.
The observations occur at times $1,\dots, N$.
{The error is measured in terms of the RMSE for the 
  mean (top row) and covariance (bottom row), computed with $N=100, 200$ and
  $400$ observation times in the first, second and third column, respectively.
  The computational cost is measured in computer runtime.} \cite{mlenkf}}
  \label{fig:mlenkf}
\end{figure}

A sufficient assumption in this case is given by 
\begin{ass}[MLEnKF] 
The coefficients of \eqref{eq:sde} are globally Lipschitz and the 
initial condition is in $L^p$ for all $p \geq 2$.
\end{ass}
The quantities for which rates $\alpha$ and $\beta$ 
need to be obtained 
are given in table \ref{tab:mlenkf},
and the complexity results are illustrated in Figure \ref{fig:mlenkf} for
an example linear SDE of the form in \eqref{eq:sde}.
The work \cite{mlenkf} established that 
slightly modified choices of $L$ and $N_{1:L}$
provide MSE at step $k$ of $\cO(|\log\epsilon|^{2n} \epsilon^{2})$ for a cost
of $\cO(\epsilon^{-2}\tilde{K}_L^{3/2})$, 
where $\tilde{K}_L^{1/2} = \sum_{l=1}^L h_l^{(\beta-\zeta)/3}$.  
However, the numerical results indicate not only 
a time-independent rate of convergence without logarithmic penalty, 
but in fact also a time-uniform constant -- 
see Figure \ref{fig:mlenkf}. 
Presumably, the penalty on the MSE is mostly a technical hurdle.
The recent work \cite{mlenkf2} has extended this method to spatial processes, 
for example given by stochastic partial differential equations.
This is the context where the EnKF is typically applied, for example 
in numerical weather prediction.

\begin{table}[h]
\begin{center}
  \begin{tabular}{ | c || c | c |}
    \hline
    Rate parameter & Relevant quantity  \\ 
    \hline\hline
    $\alpha$ & $(\bbE_{\widehat \pi^k_L} - \bbE_{\pi^k})(\varphi)$  \\ 
    \hline
    $\beta$ & 
    $\left(\int_{\bbR^{2d}} | \varphi(\overline u_k) - \varphi(\underline u_k)|^p 
    \check{\widehat{\pi}}_{l-1:l}^k(\overline u_k, \underline u_k)
    d\overline u_k d\underline u_k\right)^{2/p}$ \\ 
    \hline
  \end{tabular}
\end{center}
\caption{The key rates of convergence required for MLEnKF, for
all $p\geq 1$, where $\check{\widehat{\pi}}_{l-1:l}^k$ 
denotes the coupled measure resulting from the algorithm above.}
\label{tab:mlenkf}
\end{table}

\subsubsection{Discussion}

Some examples of  coupling algorithms have been reviewed. 
The main issues with such techniques are 
(i) 
coupling the algorithms correctly, 
so that the coupling is `good enough', and 
(ii) 
mathematical analysis of such couplings to prove that they 
indeed provide a benefit. 
These challenges are crucial and must be further studied.
It was already mentioned that the works \cite{jacob,coup_pf} 
consider coupling the pair of particle filters arising in a PMCMC algorithm,
and empirical results are promising.
However, establishing that indeed (i)-(ii) would occur in practice
is not so easy and at least does not appear to have been done in publicly available research.
In the MLMC context, the theoretical and numerical results of \cite{jasra} indicate a loss
of a power of $1/2$ in the rate of strong convergence following from the coupled resampling  
of section \ref{sssec:mlpf}, hindering the ultimate cost of the algorithm (see also table \ref{tab:mlpf}).  
However, the rates may be improved by the resampling based on optimal transportation 
from \cite{jacob,coup_pf}.  
Indeed the works \cite{reich, greg} numerically observe preservation of the strong rate of convergence 
using a deterministic transformation based on the optimal transportation coupling, 
in lieu of resampling.

\section{Future Work and Summary}\label{sec:future}

Here we examined some 
computational approaches 
to facilitate the application of the MLMC method
in challenging examples, where standard (independent)
sampling is not currently possible.
Some review of the computational methods was provided, 
although as we have noted it is a large literature that one cannot hope
to include a complete summary of all the methodology. 
We then detailed various 
approaches one can use to leverage MLMC within these methods.

There are many areas for possible exploration in future work. 
One strand consists of considering multi-dimension discretizations, such as in \cite{mimc}.
There are a small number of papers on this topic, such as \cite{ub_mimc, ourmimcmc},
but there seem to be many possible avenues for future work. 
Another direction consists of a general method for sampling (e.g.,~by MCMC/SMC) exact (dependent) couplings of the targets in the ML identity. As we have commented, it does not
appear to be trivial, but it may be far from impossible. 
Such a method would be very beneficial, as one could then
appeal to existing literature in order to prove complexity results 
about MLMC and MIMC versions.
One final very interesting avenue for future research is exact coupling, using optimal
transport and i.i.d.~sampling. For some model structures, the ideas of \cite{span} could be very useful.


\subsubsection*{Acknowledgements}

AJ and CS 
were supported under the
KAUST Competitive Research Grants Program-Round 4 (CRG4) project, 
``Advanced Multi-Level sampling techniques for Bayesian Inverse Problems with applications to subsurface.''
KJHL was supported by  
ORNL LDRD Seed funding grant number 32102582.

\end{document}